\documentclass[twocolumn,superscriptaddress,showpacs,pra,amsmath,amssymb]{revtex4}
\usepackage{graphicx}
\usepackage{dcolumn}
\usepackage{bm}
\usepackage{xcolor,hyperref}

\newcommand{\ket}[1]{|#1\rangle}
\newcommand{\bra}[1]{\langle#1|}

\newcommand{\eq}{\begin{equation}}
\newcommand{\fine}{\end{equation}}
\newcommand{\el}{\ell}
\newcommand{\kk}{\bf k}
\newcommand{\cluster}{\ket{\Phi^{\text{lin}}_4}}
\newcommand{\clthree}{\ket{\Phi^{\text{lin}}_3}}
\newcommand{\hs}{\ket{\Phi^\subset_4}}
\newcommand{\hsrot}{\ket{\Phi^\supset_4}}
\newcommand{\boxcluster}{\ket{\Phi^\Box_4}}
\begin{document}


\title{
One-way quantum computation with two-photon multiqubit cluster states
}
\author{Giuseppe Vallone}
\homepage{http://quantumoptics.phys.uniroma1.it/}
\affiliation{
Dipartimento di Fisica dell'Universit\'{a} Sapienza di Roma and\\
Consorzio Nazionale Interuniversitario per le Scienze Fisiche della Materia,
Roma, 00185 Italy}
\author{Enrico Pomarico}
\altaffiliation[Present address: ]{Univ. de Gen\`eve
GAP-Optique,
Rue de l'\'Ecole-de-M\`edecine 20,
CH-1211 Gen\`eve 4,
Suisse}
\affiliation{
Dipartimento di Fisica dell'Universit\'{a} Sapienza di Roma and\\
Consorzio Nazionale Interuniversitario per le Scienze Fisiche della Materia,
Roma, 00185 Italy}
\author{Francesco De Martini}
\homepage{http://quantumoptics.phys.uniroma1.it/}
\affiliation{
Dipartimento di Fisica dell'Universit\'{a} Sapienza di Roma and\\
Consorzio Nazionale Interuniversitario per le Scienze Fisiche della Materia,
Roma, 00185 Italy}
\affiliation{
Accademia Nazionale dei Lincei, via della Lungara 10, Roma 00165, Italy
}
\author{Paolo Mataloni}
\homepage{http://quantumoptics.phys.uniroma1.it/}
\affiliation{
Dipartimento di Fisica dell'Universit\'{a} Sapienza di Roma and\\
Consorzio Nazionale Interuniversitario per le Scienze Fisiche della Materia,
Roma, 00185 Italy}

\date{\today}

\begin{abstract}
We describe in detail the application of four qubit cluster states, built on the simultaneous entanglement of two photons in the degrees of freedom 
of polarization and linear momentum, for the realization of a complete set of basic one-way quantum computation operations.
These consist of arbitrary single qubit rotations, either probabilistic or deterministic, and simple two qubit gates,
such as a {\sc c-not} gate for equatorial qubits and a universal {\sc c-phase} (CZ) gate acting on arbitrary
target qubits.
Other basic computation operations, such as the Grover's 
search and the Deutsch's algorithms, have been realized by using these states. In all the cases we obtained a high value of the operation fidelities. 
These results demonstrate that cluster states of two photons entangled in many degrees of freedom are good candidates for the realization of more 
complex quantum computation operations based on a larger number of qubits.

\end{abstract}

\maketitle

\section{Introduction}\label{sec:intro}
The relevance of cluster states in quantum information and quantum computation (QC) has been emphasized 
in several papers in recent years \cite{01-rau-aon,04-nie-opt,05-bro-res,05-kie-exp,05-wal-exp,06-bod-sca,06-dan-det,07-pre-hig,07-pop-kni,08-var-how,07-lu-exp}.
By these states novel significant tests of quantum nonlocality, 
which are more resistant to noise and show significantly larger deviations from classical bounds can be realized
\cite{05-cab-str,05-sca-non,07-val-rea,05-kie-exp}.  

Besides that, cluster states represent today the basic resource for the realization of a quantum computer operating 
in the one-way model \cite{01-rau-aon}. 
In the standard QC approach any quantum algorithm can be realized by a sequence of single qubit rotations and two qubit gates, 
such as {\sc c-not} and {\sc c-phase} on the physical qubits \cite{00-nie-qua,01-kni-asc,07-kok-lin}, while deterministic one-way QC is based on
the initial preparation of the physical qubits in a cluster state, followed by
a temporally ordered pattern of single qubit measurements and feed-forward (FF)
operations \cite{01-rau-aon}.
By exploiting the correlations existing between the physical qubits,
the unitary gates on the so called ``encoded'' (or logical) qubit \cite{05-wal-exp} are realized.
In this way, non-unitary measurements on the physical qubits correspond to unitary gates on the logical qubits.
It is precisely this non-unitarity of the physical process that causes the irreversibility nature (i.e. its ``one-way'' feature) of the model. 
By this model the difficulties of standard QC, related to the implementation of two qubit gates, 
are transferred to the preparation of the state.

The FF operations, that depend on the outcomes of the already
measured qubits and are necessary for a deterministic computation, can be classified in two classes:
\begin{itemize}
 \item[\it i)] the intermediate \textit{feed-forward measurements}, i.e. the choice of the measurement basis;
\item[\it ii)] the Pauli matrix \textit{feed-forward corrections} on the
final output state. 
\end{itemize}

The first experimental results of one-way QC, either probabilistic or deterministic, were demonstrated by using 4-photon cluster states
generated via post-selection by spontaneous parametric down conversion (SPDC).
\cite{05-wal-exp,07-pre-hig}. 
The detection rate in such experiments, approximately 1 Hz, was limited by the fact that four photon events 
in a standard SPDC process are rare. 
Moreover, four-photon cluster states are characterized by limited values of fidelity, while 
efficient computation requires the preparation of highly faithful states.

More recently, by entangling
two photons in more degrees of freedom,
we created four-qubit cluster states at a much higher level of brightness and fidelity \cite{07-val-rea}. 
Precisely, this was demonstrated by entangling the polarization ($\pi$) and linear momentum ({\bf k})
degrees of freedom of one of the two photons belonging to a hyperentangled state \cite{05-bar-pol,06-bar-enh}.
Moreover, working with only two photons allows to reduce the problems related to the limited quantum efficiency
of detectors.  
Because of these characteristics, two-photon four-qubit cluster states 
are suitable for the realization of high speed one-way QC \cite{08-val-act, 08-val-one, 07-che-exp}.

In this paper we give a detailed description of the basic QC operations performed by using two-photon four-qubit 
cluster state, such as arbitrary single qubit rotations, the {\sc c-not} gate for equatorial qubits and a {\sc c-phase} gate.
We verified also the equivalence existing 
between the two degrees of freedom for qubit rotations, by using either $\bf k$ or $\pi$ as output qubit, 
demonstrating that all four qubits can be adopted for computational applications. 
Moreover, we also show the realization of two important basic algorithms by our setup, 
namely the Grover's search algorithm and the Deutsch's algorithm. 
The former identifies the item tagged by a ``Black Box", while the latter allows to distinguish
in a single run if a function is {\it constant} or {\it balanced}.

The paper is organized as follows. In Sec. \ref{sec:one-way} we review the one-way model of QC realized through single qubit
measurements on a cluster state. We also describe the basic building blocks that can be used in order to implement any general algorithm.
In Sec. \ref{sec:experiment} we give a description of the source used to generate the two-photon four-qubit cluster state
by manipulating a polarization-momentum hyperentangled state. We describe in Sec. \ref{sec:operations} three basic operations
realized by our setup: a generic single qubit rotation, a {\sc c-not} gate for equatorial target qubit and a {\sc c-phase}
gate for fixed control and arbitrary target qubit. In Sec. \ref{sec:algorithms} two explicit examples of quantum computation are given
by the realization of the Grover's search algorithm and the the Deutsch's algorithm.
Finally, the conclusions are given in Sec. \ref{sec:conclusions}.

\section{One-way computation}\label{sec:one-way}
\begin{figure}
  \centering
    \includegraphics[width=8cm]{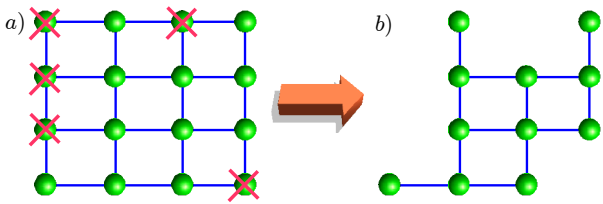}
  \caption{(color online). Effect of measurements on a generic cluster state. a) Measurements in the computational basis $\{\ket0,\ket1\}$ are indicated by red crosses. 
        b) The cluster state after the measurements of some qubits in the computational basis.}
  \label{fig:lattice}
\end{figure} 
Cluster states are quantum states associated to $n$ dimensional lattices.
It has been shown that two-dimensional cluster states are a universal resource for QC \cite{05-nie-clu}.
The explicit expression of a cluster state is found by associating to each dot $j$ of the lattice
(see fig. \ref{fig:lattice}) a qubit in the state ${\ket{+}}_j=\frac1{\sqrt2}({\ket0}_j+{\ket1}_j)$ and to each link 
between two adjacent qubits $i$ and $j$, a {\sc c-phase} gate $CZ_{ij}$:
\eq
CZ_{ij}={\ket0}_i\bra0\otimes\openone_j+{\ket1}_i\bra1\otimes\sigma^{(j)}_z\,.
\fine 
Considering a lattice $\mathcal L$ with $N$ sites, the corresponding cluster state is then given by the expression:
\eq
\ket{\Phi^{\mathcal L}_N}\equiv\left(\prod_{i,j\text{ linked}}CZ_{ij}\right)\ket+^N\,,
\fine
where $\ket+^N\equiv\ket{+}_1\otimes\ket+_2\cdots\otimes\ket{+}_N$.
Some explicit examples of cluster states are the 3-qubit linear cluster,
\begin{equation}
\clthree=\frac1{\sqrt2}(\ket{+}_1\ket{0}_2\ket{+}_3+\ket{-}_1\ket{1}_2\ket{-}_3)\,,
\end{equation}
the 4-qubit linear (or horseshoe) cluster
\eq
\begin{aligned}\label{4qubit-cluster}
\cluster=&\hs=\hsrot=\\
=&\frac12(\ket{+}_1\ket{0}_2\ket{0}_3\ket{+}_4+\ket{+}_1\ket{0}_2\ket{1}_3\ket{-}_4\\
&+\ket{-}_1\ket{1}_2\ket{0}_3\ket{+}_4
+\ket{-}_1\ket{1}_2\ket{1}_3\ket{-}_4)\,,
\end{aligned}
\fine
corresponding to four qubits linked in a row (see Fig. \ref{fig:rotation}I))
and the 4-qubit Box cluster
\begin{multline}
\boxcluster=\frac12(\ket{0}_1\ket{+}_2\ket{0}_3\ket{+}_4
+\ket{0}_1\ket{-}_2\ket{1}_3\ket{-}_4
\\
+\ket{1}_1\ket{-}_2\ket{0}_3\ket{-}_4
+\ket{1}_1\ket{+}_2\ket{1}_3\ket{+}_4)\,.
\end{multline}
corresponding to four qubits linked in a square (see Fig. \ref{fig:grover}(left)).

For a given cluster state, the measurement of a generic qubit $j$ performed in the computational 
basis $\{{\ket0}_j,{\ket1}_j\}$ (Fig. \ref{fig:lattice}a)) 
simply corresponds to remove it and its relative links from the cluster (Fig. \ref{fig:lattice}b)).
In this way we obtain, up to possible $\sigma_z$ corrections, a cluster state with $N-1$ qubits:
\eq
\ket{\Phi^{\mathcal L}_N}\rightarrow
\prod_{k\in \mathcal N_j}(\sigma^{(k)}_z)^{s_j}\ket{\Phi^{\mathcal L\backslash\{j\}}_{N-1}}\,,
\fine
where $s_j=0$ if the measurement output is ${\ket0}_j$, while
$s_j=1$ if the measurement output is ${\ket1}_j$. In the previous equation $\mathcal N_j$ stands for the set of all
sites linked with the qubit $j$.
Then, by starting from a large enough square lattice, it is possible to create any cluster state associated
to smaller lattices. In the following figures we will indicate with a red cross the measurement of a physical qubit 
performed in the computational basis, as shown in Fig. \ref{fig:lattice}.

Let's now explain how the computation proceeds. 
Each algorithm consists of a measurement pattern on a specific cluster state. This pattern has a precise temporal ordering. 
It is well known that one-way computation doesn't operate directly on the physical qubits 
of the cluster state on which measurements are performed. 
The actual computation takes place on the so-called encoded qubits, written nonlocally in the cluster through the correlation
between physical qubits. We will use $i,j=1,\cdots, N$ for the physical qubits and $a,b=1,\cdots, M$ for the encoded qubits ($M<N$).
Some physical qubits (precisely $M$) represent the input qubits of the computation (all prepared in the state $\ket{+}_E$)
and the corresponding dots can be arranged at the left of the graph.
We then measure $N-M$ qubits, leaving $M$ physical qubits unmeasured, hence the output of computation 
will correspond (up to Pauli errors) to the unmeasured qubits.
It's possible to arrange the position of the dots in such a way that the time ordering 
of the measurement pattern goes from left to right.  

The computation proceeds by the measurement performed in the basis 
\eq
B_j(\varphi)\equiv\{{\ket{\varphi_+}}_j,{\ket{\varphi_-}}_j\}\,,
\fine 
where
${\ket{\varphi_\pm}}_j\equiv\frac1{\sqrt2}(e^{i\varphi/2}{\ket0}_j\pm e^{-i\varphi/2}{\ket1}_j)$. Here 
$s_j=0$ or $s_j=1$ if the measurement outcome of qubit $j$ is ${\ket{\varphi_+}}_j$ or ${\ket{\varphi_-}}_j$  respectively.
The specific choice of $\varphi$ for every physical qubit is determined by the measurement pattern. Note
that the choice of the measurement basis for a specific qubit can also depend on the outcome of the already measured qubits:
these are what we call feed-forward measurements (type \textit{i)}).
In general, active modulators (as for example Pockels cells in case of polarization qubit) are required to perform the FF measurements. 
In some case, however, when more than one qubit is encoded in the same particle through different degrees of freedom (DOF's), 
it is possible to perform FF measurement without the use of active modulators. 
This will be discussed in Sec. \ref{sec:operations}, when the measurement basis of the generic qubit $j$, encoded in one particle, 
depends only on the outputs of some other qubits encoded in the same particle.

\begin{figure}
  \centering
    \includegraphics[width=8cm]{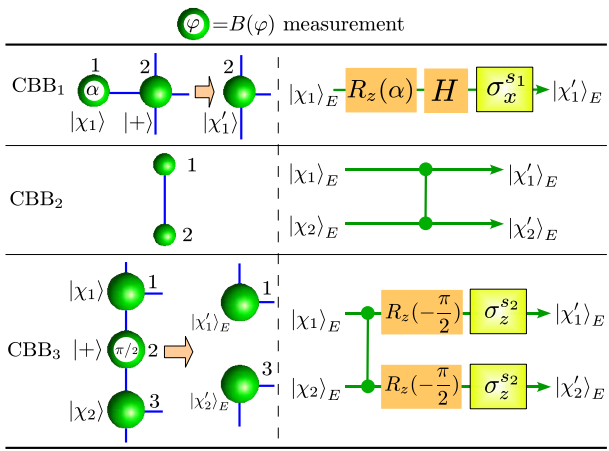}
  \caption{(color online). Cluster building blocks (CBB). For each CBB$_i$ we indicate the measurement on the physical cluster 
          (left) and the corresponding operation on the logical qubits (right).}
  \label{fig:CBB}
\end{figure}

One-way computation can be understood in terms of some basic operations, the so-called cluster building blocks (CBB) (see Fig. \ref{fig:CBB}).
By combining different CBB's it becomes possible to perform computation of arbitrary complexity \cite{05-tam-qua}.
We introduce here a convenient notation: by writing explicitely a state $\ket\chi$ close to a dot $j$ (see Fig. \ref{fig:CBB}),
we indicate that the total state could be equivalently obtained by preparing the
qubit $j$ in the state $\ket\chi_j$ before applying the necessary $CZ$ gates.

{\bf CBB$_1$: Qubit rotation}. 
Consider two qubits linked together like CBB$_1$ shown in Fig. \ref{fig:CBB}. Here the first qubit is
initially prepared in the state $\ket{\chi}$ and the second qubit is arbitrary linked with other dots.
By measuring qubit 1 in the basis $B_1(\alpha)$ we remove it from the cluster but we 
transfer the information into qubit $2$ leaving its links unaltered.
This corresponds to the following operation on the encoded qubit $\ket{\chi}_E$
\eq
{\ket{\chi}}_E\rightarrow{\ket{\chi'}}_E\equiv\sigma_x^{s_1}HR_z(\alpha){\ket{\chi}}_E\,,
\fine
where $H$ is the Hadamard gate $H=1/\sqrt2(\sigma_x+\sigma_z)$ and $R_z(\alpha)=e^{-i\alpha\sigma_z/2}$ is
a rotation around the $z$ axis in the Bloch sphere. The $\sigma_x$ operations depends on the measurement output ($s_1$) of the 
first qubit.
This operation can be understood by noting that by measuring the first qubit of the state
$CZ_{12}{\ket{\chi}}_1{\ket+}_2$ in the basis $B_1(\alpha)$ we will obtain 
${\ket{\chi'}}_2$. 

This simple algorithm can be repeated by 
using two CBB$_1$ in a row.
By measuring qubit 1 in the bases $B_1(\alpha)$ the encoded qubit then is transformed into $\ket{\chi'}_E$ and
the encoded qubit moves from left to right into the cluster.
By measuring qubit 2 in the basis $B_2(\beta)$ the encoded qubit is now written in qubit 3 as $\ket{\chi''}_E$:
\eq
{\ket{\chi''}}_E\equiv\sigma_x^{s_2}HR_z(\beta)\sigma_x^{s_1}HR_z(\alpha){\ket{\chi}}_E\,.
\fine
In this case the computation can be understood by observing that with the measurement of 
qubits $1$ and $2$ of the state $CZ_{12}CZ_{23}{\ket{\chi}}_1{\ket+}_2{\ket+}_3$ in the basis 
$B_1(\alpha)$ and $B_2(\beta)$, we obtain $\ket{\chi''}_3$.

{\bf CBB$_2$: {\sc c-phase} gate}. 
Consider two qubits linked in a column.
This block simply performs a {\sc c-phase} gate ($CZ$) between the two qubits.

{\bf CBB$_3$: {\sc c-phase} gate+rotation}. If we have 3 qubits in a column and we measure the second qubit in the basis $B_2(\pi/2)$ we 
remove it from the cluster but the information is transferred to the qubit 1 and 3.
Precisely, on the logical qubit, this measurement realizes a {\sc c-phase} gate followed by two single qubit rotations $R_z(-\pi/2)$ (see Fig. \ref{fig:CBB}).

By combining these CBB's we can obtain any desired quantum algorithm, written in general as
\eq
\ket{\psi_{out}}=U_\Sigma U_g\prod^M_{a=1}\ket{+}_a\,,
\fine
where $M$ are the number of logical qubits, $U_g$ is the unitary gate that the algorithm must perform and
$U_\Sigma$ are the so called Pauli errors corrections \cite{03-rau-mea,05-tam-qua}: 
\eq
U_\Sigma=\prod^M_{a=1}(\sigma^{[a]}_x)^{x_a}(\sigma^{[a]}_z)^{z_a}\,.
\fine
The numbers $x_a,z_a=0,1$ depend on the outcomes of all the single (physical) qubit measurements and
determine the FF corrections (type \textit{ii)}) that must be realized, at the end of the measurement pattern, in order to achieve
deterministic computation. We indicate by the symbol $\sigma^{[a]}_z$ that
the Pauli matrix $\sigma_z$ acts on the logical qubit $a$. Note that if the output of the algorithm is one among the $2^M$ states of the computational
basis, i.e. $\bigotimes^M_{a=1}{\ket{r_a}}_a$ ($r_a=0,1$), only the $\sigma_x$'s of the unitary $U_\Sigma$ act non-trivially by flipping some qubits.
The Pauli errors are then reduced to
\eq\label{pauli_error}
U'_\Sigma=\prod^M_{a=1}(\sigma^{[a]}_x)^{x_a}\,.
\fine
In this way the ``errors'' can be simply corrected by relabeling the output, and there is no need of active feed-forward corrections on
the quantum state.
If, by measuring the output qubits, we get the result $\prod^M_{a=1}{\ket{s_a}}_a$ ($s_a=0$ or $s_a=1$) 
we must interpret it as $\prod^M_{a=1}{\ket{s_a\oplus x_a}}_a$ with the Pauli errors given by the equation \eqref{pauli_error}.
This relabeling operation can be performed for example by an external classical computer.

\section{Experimental setup}\label{sec:experiment}
\begin{figure}[t]
	\begin{center}
		\includegraphics[width=8cm]{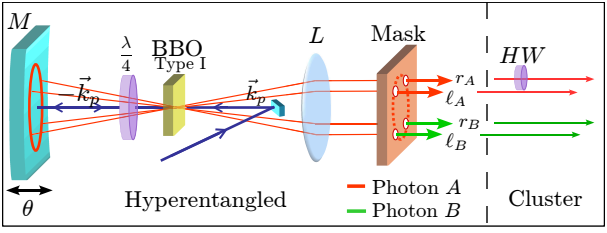}
	\end{center}
		\caption{(color online). Source of two-photon four-qubit cluster state. The hyperentangled states $\ket{\Xi^{\pm\pm}}$ 
                derive from the simultaneous entanglement on the polarization and linear momentum degrees of freedom.
                Polarization entanglement is obtained by double passage of the pump and SPDC pair through a BBO type I 
                crystal and a lambda/4 wave plate. Mode selection performed by a four hole screen allows linear momentum 
                entanglement. The half wave plate $HW$ transforms the hyperentangled state $\ket{\Xi^{+-}}$ into the 
                cluster state $\ket{C_4}$.}
\label{fig:source}
\end{figure}
In our experiment we generated two-photon four-qubit cluster states by starting from polarization($\pi$)-momentum($\bf k$)
hyperentangled photon pairs obtained by SPDC (see Fig. \ref{fig:source}).
The hyperentangled states 
$\ket{\Xi^{\pm\pm}}\equiv\ket{\Phi^\pm}_\pi\otimes\ket{{\psi^\pm}}_{\bf k}$ were generated by a  
$\beta$-barium-borate (BBO) type I crystal pumped in both sides by a cw Argon laser beam ($\lambda_p=364nm$) (cfr. \ref{fig:source}).
The detailed explanation of hyperentangled state preparation was given in previous papers \cite{05-bar-pol,05-cin-all}, 
to which we refer for details.
In the above expression of $\ket{\Xi^{\pm\pm}}$ we used the Bell states
\eq
\begin{aligned}
&|\Phi^\pm\rangle_\pi =\frac{1}{\sqrt{2}}
\left( |H\rangle _{A}|H\rangle_{B}\pm|V\rangle _{A}|V\rangle _{B}\right)\,,\\
&|{\psi^\pm} \rangle_{\bf k} =\frac{1}{\sqrt{2}}(|\ell \rangle_{A}|r\rangle _{B}\pm|r\rangle _{A}|\ell \rangle _{B}),
\end{aligned}
\fine
where 
$\ket H,\ket V$ correspond to the horizontal ($H$) and
vertical ($V$) polarizations and $\ket \ell ,\ket r$ refer to the left ($\ell $) or
right ($r$) paths of the photon $A$ (Alice) or $B$ (Bob) (see Fig. \ref{fig:source}).
In $\ket{\Xi^{\pm\pm}}$, the first signs refer to the polarization state $\ket{\phi^\pm}_\pi$
and the second ones to the linear momentum state $\ket{\psi^\pm}_{\bf k}$.
\begin{ruledtabular}
\begin{table}
\centering
 \begin{tabular}{rcc}
  &  Stabilizers &  Tr[$\rho_{exp}S_k$]  \\
\hline
$S_1$& $-z_Az_B$ & $0.9941\pm0.0011$
\\
$S_2$& $-x_Ax_BZ_A$ & $0.8486 \pm 0.0031$
\\
$S_3$& $z_AX_AX_B$ & $0.9372 \pm 0.0035$
\\
$S_4$& $Z_AZ_B$ & $0.9105\pm 0.0024$
\\
$S_5$& $-y_Ay_BZ_A$ & $ 0.8386\pm0.0032$
\\
$S_6$& $-z_BX_AX_B$ & $ 0.9354 \pm 0.0035$
\\
$S_7$& $-Z_AZ_Bz_Az_B$ & $ 0.8963 \pm 0.0044$
\\
$S_8$& $Y_AX_Bx_Ay_B$ & $ 0.7455 \pm 0.0042$
\\
$S_9$& $-Z_By_Ay_B$ & $ 0.8215 \pm 0.0034$
\\
$S_{10}$&$ X_AY_Bx_AY_B$ & $ 0.8139 \pm 0.0037$
\\
$S_{11}$& $-Y_AX_By_Ax_B$ & $ 0.7944 \pm 0.0037$
\\
$S_{12}$& $-x_Ax_BZ_B$ & $ 0.8498 \pm 0.0031$
\\
$S_{13}$& $-Y_AY_Bz_A$ & $ 0.9350 \pm 0.0036$
\\
$S_{14}$& $Y_AY_Bz_B$ & $ 0.9346 \pm 0.0037$
\\
$S_{15}$& $-X_AY_By_Ax_B$ & $ 0.8186 \pm 0.0035$
\\
$S_{16}$& $\openone$ & $1$
 \end{tabular}
 \caption{Expectation values of the stabilizer operators $S_i$.}
  \label{table:stabilizer}
\end{table}
\end{ruledtabular}

Starting from the state $\ket{\Xi^{+-}}=\ket{\Phi^+}_{\pi}\otimes\ket{\psi^-}_{\bf k}$ and applying
a {\sc c-phase} (CZ) gate 
between the control ($\bf k_A$) and the target ($\pi_A$) qubits of the photon A,
we generated the cluster state
\eq
\begin{aligned}\label{cluster}
\ket{C_{4}}=&\frac{1}{2}(|H\ell \rangle _{A}|Hr\rangle _{B}- |Hr\rangle _{A}|H\ell \rangle_{B}\\
&+|Vr\rangle _{A}|V\ell \rangle _{B}+|V\ell \rangle _{A}|Vr\rangle_{B})\\
=&\frac1{\sqrt2}{\ket{\Phi^+}}_\pi{\ket{\ell}}_A{\ket{r}}_A-\frac1{\sqrt2}{\ket{\Phi^-}}_\pi{\ket{r}}_A{\ket{\ell}}_A\\
=&\frac1{\sqrt2}{{\ket{H}}_A{\ket{H}}_A\ket{\psi^+}}_{\kk}+\frac1{\sqrt2}{{\ket{V}}_A{\ket{V}}_A\ket{\psi^-}}_{\kk}\,.
\end{aligned}
\fine
In the experiment, the CZ gate is realized by a zero-order half wave (HW) plate inserted on the $r_A$ mode,
and corresponds to introduce a $\pi$ phase shift for the vertical polarization of the $r_A$ output mode.
It is worth noting that, at variance with the four-photon cluster state generation, the state \eqref{cluster} is created without any kind of 
post-selection\footnote{Note that the usual post-selection needed to select out the vacuum state is in any case necessary. 
This is unavoidable since SPDC is a not deterministic process.}.
By using the correspondence $\ket H\leftrightarrow\ket0$, $\ket V\leftrightarrow\ket1$,
$\ket \el\leftrightarrow\ket0$, $\ket r\leftrightarrow\ket1$,
the generated state $\ket{C_4}$ is equivalent to $\cluster$, $\hs$, 
$\hsrot$ or $\boxcluster$ 
up to single qubit unitaries:
\eq\label{equiv cluster}
\ket{C_4}=U_1\otimes U_2\otimes U_3\otimes U_4\cluster\equiv\mathcal U\cluster\,.
\fine
With $\cluster$ and $\ket{C_4}$ we will refer to the cluster state expressed in the ``cluster'' and ``laboratory'' basis respectively. 
The explicit expression of the unitaries $U_j$ depends on the specific ordering of the physical qubits ($\bf k_A$, $\bf k_B$, $\pi_A$ and $\pi_B$)
and will be indicated in each case in the following.
The basis changing will be necessary in order to know which are the correct measurements we need to perform in the actual experiment.
In general if on qubit $j$ the chosen algorithm requires a measurement in the basis ${\ket{\alpha_\pm}}_j$, the actual
measurement basis in the laboratory is given by $U_j{\ket{\alpha_\pm}}_j$.

In order to characterize the generated state we used the stabilizer operator formalism \cite{97-got-sta}.
It can be shown \cite{06-hei-ent} that
\eq
\ket{C_4}\bra{C_4}=\frac1{16}\sum^{16}_{k=1}S_k\,,
\fine
where the $S_k$ are the so called {\it stabilizer operators} $S_k\ket{C_4}=\ket{C_4}$, $\forall k=1,\cdots,16$ (see table \ref{table:stabilizer}).
The fidelity of the experimental cluster $\rho_{exp}$ can be measured by 
\eq
F_{\ket{C_4}}=\text{Tr}[\rho_{exp}\ket{C_4}\bra{C_4}]=
\frac1{16}\sum^{16}_{k=1}\text{Tr}[\rho_{exp}S_k]\,,
\fine
i.e. by measuring the expectation values of the stabilizer operators. In table \ref{table:stabilizer} we
report the stabilizer operators for $\ket{C_4}$ and the corresponding experimental expectation values.
The obtained fidelity is $F = 0.880\pm0.013$, demonstrating the high purity level of the generated state.
Cluster states were observed at 1 kHz detection rate. The two photons are detected by using interference filter with bandwidth $\Delta \lambda=6nm$.
In table \ref{table:stabilizer} we use the following notation for polarization
\eq\label{pol}
\begin{aligned}
&Z_j=\ket{H}_j\bra{H}-\ket{V}_j\bra{V}\\
&Y_j=i\ket{V}_j\bra{H}-i\ket{H}_j\bra{V}\\
&X_j=\ket{H}_j\bra{V}+\ket{V}_j\bra{H}
\end{aligned}
\qquad \qquad j=A,B
\fine
and linear momentum operators
\eq\label{mom}
\begin{aligned}
&z_j=\ket{\ell}_j\bra{\ell}-\ket{r}_j\bra{r}\\
&y_j=i\ket{r}_j\bra{\ell}-i\ket{\ell}_j\bra{r}\\
&x_j=\ket{\ell}_j\bra{r}+\ket{r}_j\bra{\ell}
\end{aligned}
\qquad \qquad j=A,B
\fine
for either Alice (A) or Bob (B) photons.

\section{Basic operations with 2-photon cluster state}\label{sec:operations}
In this section we describe the implementation of simple operations performed with the generated four-qubit two-photon cluster state.
\subsection{Single qubit rotations}\label{subsec:rotations}
\begin{figure}[t]
	\begin{center}\
		\includegraphics[width=7cm]{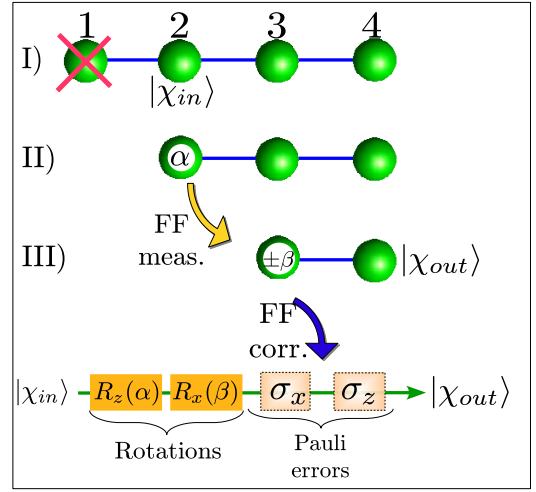}
	\end{center}
		\caption{(color online). Measurement pattern for single qubit rotations. (a) Top: arbitrary single qubit rotations on a four
		qubit linear cluster state are carried out in three steps (I, II, III). In each measurement, indicated by a red cross, 
		the information travels from left to right. Bottom: equivalent logical circuit. }
\label{fig:rotation}
\end{figure}

In the one-way model a three-qubit linear cluster state (simply obtained by the four-qubit cluster by measuring the first qubit)
is sufficient to realize an arbitrary single qubit transformation\footnote{This is a generic rotation iff the input state $\ket{\chi_{in}}$ is not $\ket0$ or $\ket1$.
In our case the algorithm is implemented with $\ket{\chi_{in}}=\ket\pm$.
In fact three sequential rotations are in general necessary to implement a generic $SU(2)$ matrix but only two, 
namely $R_x(\beta)R_z(\alpha)$, are sufficient to rotate the input state $\ket{\chi_{in}}=\ket\pm$ into a generic state}
$\ket{\chi_{in}}\rightarrow R_x(\beta)R_z(\alpha)\ket{\chi_{in}}$,
where $R_z(\alpha)=e^{-i\alpha\sigma_z/2}$ and $R_x(\beta)=e^{-i\beta\sigma_x/2}$.


The algorithm consists of two CBB$_1$'s on a row. Precisely, by using the four-qubit cluster expressed in the cluster basis 
the following measurement pattern must be followed (see Fig. \ref{fig:rotation}):
\begin{itemize}
\item[I:] A three-qubit linear cluster is generated by measuring the first qubit in the  computational basis $\{{\ket0}_1, {\ket1}_1\}$. 
As said, in Section \ref{sec:one-way} this operation removes the first qubit from the cluster and generates 
$(\sigma^{(2)}_z)^{s_1}\ket{\Phi^{\text{lin}}_3}$.
The input logical qubit $\ket{\chi_{in}}$ is then encoded in qubit 2. If the outcome of the first measurement is 
${\ket0}_1$ then $\ket{\chi_{in}} = \ket+$, otherwise $\ket{\chi_{in}}=\ket-$.
\item[II:] Measuring qubit 2 in the basis $B_2(\alpha)$, the logical qubit (now encoded in qubit 3) 
is transformed into
$\ket{\chi'}={(\sigma_x)}^{s_2}HR_z(\alpha)\ket{\chi_{in}}$,
with $R_z(\alpha)=e^{-\frac{i}{2}\alpha \sigma_z}$.
\item[III:] Measurement of qubit 3 is performed in the basis:
 \begin{itemize}
  \item[$\bullet$] $B_3(\beta)$ if $s_2=0$
  \item[$\bullet$] $B_3(-\beta)$ if $s_2=1$
 \end{itemize}
This represents a FF measurement (type {\it i)}) since the choice of measurement basis depends on the previous outputs.
This operation leaves the last qubit in the state 
$\ket{\chi_{out}}={(\sigma_x)}^{s_3}HR_z\left[(-1)^{s_2}\beta\right]\ket{\chi'}$
with $R_x(\beta)=e^{-\frac{i}{2}\beta \sigma_x}$.
\end{itemize}
Then, by using some simple Pauli matrix algebra, the output state (encoded in qubit $4$) can be written as
\eq
\ket{\chi_{out}}=\sigma^{s_3}_x\sigma^{s_2}_zR_x(\beta)R_z(\alpha)\ket{\chi_{in}}\,.
\fine
In this way, by suitable choosing $\alpha$ and $\beta$,
we can perform any arbitrary single qubit rotation $\ket{\chi_{in}}\rightarrow
R_x(\beta)R_z(\alpha)\ket{\chi_{in}}$
up to Pauli errors ($\sigma^{s_3}_x\sigma^{s_2}_z$), that should be corrected by proper feed-forward 
operations (type {\it ii)}) to achieve a deterministic computation\cite{07-pre-hig}. 
\begin{figure}
	\centering
	\includegraphics[width=8cm]{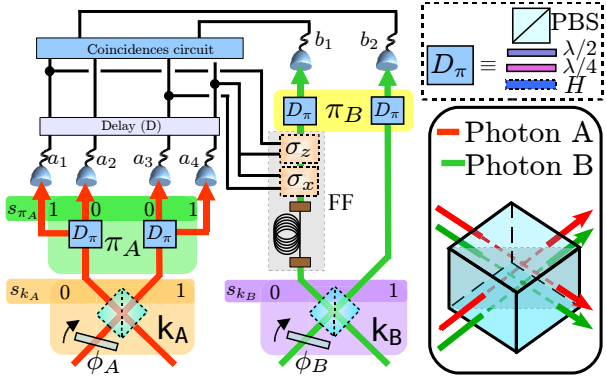}
	\caption{(color online). Measurement setup for photons A and B. Momentum qubits $\bf{k}_A$ and $\bf{k}_B$ are measured 
by two thin glass ($\phi_A$, $\phi_B$), acting as phase shifters and inserted before a common 50:50 BS. 
Polarization qubits $\pi_A$ and $\pi_B$ are measured
by standard tomographic setup ($D_\pi$).
BS and $D_\pi$ outputs are indicated by $s_j=0, 1$, where the index $j$ refers to the corresponding DOF. 
FF correction apparatus for deterministic QC [used in our experiment only with ordering (a)] is
given by a 35 m length single mode fiber and two Pockels cells
($\sigma_x$, $\sigma_z$) driven by the output signals of detector $a_1$, $a_3$, $a_4$.
Dashed lines for $H$, $BS$ and FF correction apparatus indicate
that these devices can be inserted or not in the setup 
depending on the particular measurement (see text for details). Inset: tomographic apparatus $D_\pi$ and spatial mode matching on the BS.
}
	\label{fig:schema}
\end{figure}

In our case we applied this measurement pattern by considering different ordering of the physical qubits.
In this way we encoded the output qubit either in the polarization or in the linear momentum of photon $B$, demonstrating the QC equivalence of the two degrees of freedom.
The measurement apparatus is sketched in Fig. \ref{fig:schema}. The $\bf k$ modes corresponding to photons A or B, are 
respectively matched on the up and down side of a common symmetric beam splitter (BS) (see inset), which
can be also finely adjusted in the vertical direction such that one or both photons don't pass through it.
Polarization analysis is performed by a standard tomography apparatus $D_\pi$
($\lambda/4$, $\lambda/2$ and polarizing beam splitter PBS).
Depending of the specific measurement the HWs oriented at 22.5$^\circ$ are inserted to perform the Hadamard operation H
in the apparatus $D_\pi$. They are used together with the $\lambda/4$ in order to transform the
$\{\ket{\varphi_+}_{\pi_A},\ket{\varphi_-}_{\pi_A}\}$ states into linearly polarized states.
Two thin glass plates before the BS allow to set the basis of the momentum measurement for each photon.

\begin{figure}[t]
	\begin{center}
		\includegraphics[width=8.5cm]{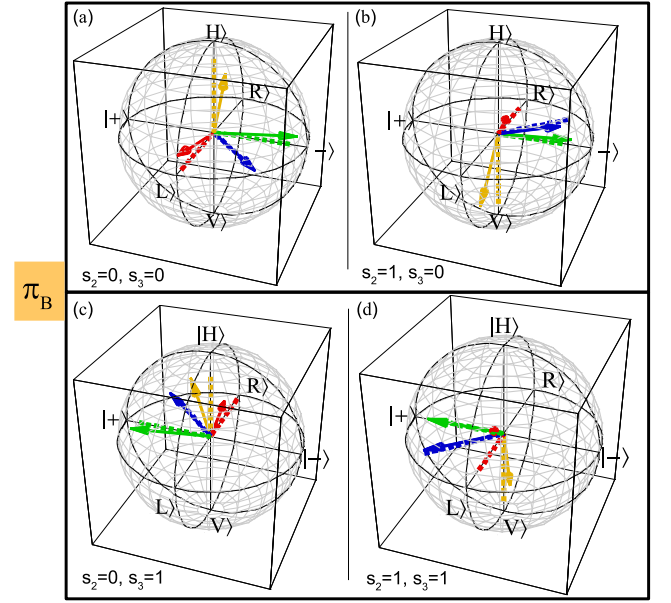}
	\end{center}
		\caption{(color online). Polarization ($\pi_B$) output Bloch vectors of single qubit rotations. 
The experimental results (arrows) are shown with their projections on theoretical directions (dashed lines). 
Arrow colours correspond to different values of $\alpha$ and $\beta$ (see table \ref{table:rotation_pi}). }
\label{fig:bloch_pi}
\end{figure}
\begin{table}
	\begin{ruledtabular}
				\centering
		\begin{tabular}{ccccc}
			&$\alpha$ & $\beta$ & F $(s_2=0,s_3=0)$ & F $(s_2=0,s_3=1)$ 
			\\
			\cline{1-5}					
			&$0$      &  $\pi/2$    & $0.908\pm0.006$ & $0.860\pm0.008$ 
			\\
$\pi_B$ &$-\pi/2$ & $0$    & $0.942\pm0.004$ & $0.946\pm0.004$ 
			\\
			&$-\pi/2$      &  $\pi/2$   & $0.913\pm0.005$ & $0.929\pm0.004$ 
			\\
			&$-\pi/2$      &  $-\pi/4$   & $0.899\pm0.007$ & $0.898\pm0.006$ 
			\end{tabular}

		\begin{tabular}{ccccc}
			&$\alpha$ & $\beta$ & F $(s_2=1,s_3=0)$ & F $(s_2=1,s_3=1)$ 
			\\
			\cline{1-5}					
			&$0$      &  $\pi/2$    & $0.932\pm0.005$ & $0.935\pm0.006$ 
			\\
$\pi_B$ &$-\pi/2$      &  $0$    & $0.873\pm0.006$ & $0.847\pm0.006$ 
			\\
			&$-\pi/2$      &  $\pi/2$   & $0.851\pm0.007$ & $0.848\pm0.007$
			\\
			&$-\pi/2$      &  $-\pi/4$   & $0.928\pm0.007$ & $0.932\pm0.007$
		\end{tabular}
\end{ruledtabular}
		\caption{(color online). Polarization ($\pi_B$) 
		experimental fidelities (F) of single qubit rotation output states for different values of $\alpha$ and $\beta$. 
		Each datum is obtained by the measurements of the different Stokes parameters, each one lasting 10 sec. }
\label{table:rotation_pi}
\end{table}

Let's consider the following ordering of the physical qubits (see Eq. \eqref{equiv cluster}):
\eq\label{order_a}
\begin{aligned}
a)\qquad \text{(1,2,3,4)=}&({\bf k}_B,{\bf k}_A,\pi_A,\pi_B), \\
\mathcal U=&\sigma_xH\otimes\sigma_z\otimes\openone\otimes H\,.
\end{aligned}
\fine
The output state, encoded in the polarization of photon B, can be written in the laboratory basis as
\eq\label{out}
|\chi _{out}\rangle_{\pi _{B}}={\sigma _{z}}^{s_{\pi_A}}{\sigma _{x}}^{s_{{\kk}_A}}HR_{x}(\beta)
R_{z}(\alpha )|\chi _{in}\rangle \,,
\fine
where the $H$ gate derives from the change between the cluster and laboratory basis.
This also implies that the actual measurement bases are $B_{\bf k_B}(0)$ for
the momentum of photon B (qubit 1) and $B_{\bf k_A}(\alpha+\pi)$ (i.e. $\ket{\alpha_{\mp}}_{\bf k_A}$) for the momentum of photon A (qubit 2).
According to the one-way model, the measurement basis on the third qubit ($\pi_A$) 
depends on the results of the measurement on the second qubit ($\bf k_A$).
These are precisely what we call FF measurements (type \textit{i)}). In our scheme this simply corresponds to measure 
$\pi _{A}$ in the bases $B_{\pi _{A}}(\beta)$ or $B_{\pi _{A}}(-\beta)$, depending on the BS
output mode (i.e. $s_{\kk_A}=0$ or $s_{\kk_A}=1$). 
These deterministic FF measurements are a direct consequence of the possibility to encode two qubits ($\bf k_A$ and $\pi_A$) in the same photon.
As a consequence, at variance with the case of four-photon cluster states,
in this case active feed-forward measurements (that can be realized by adopting Pockels cells) are not required, while Pauli
errors corrections are in any case necessary for deterministic QC.
\begin{figure}[t]
	\begin{center}
		\includegraphics[width=8.5cm]{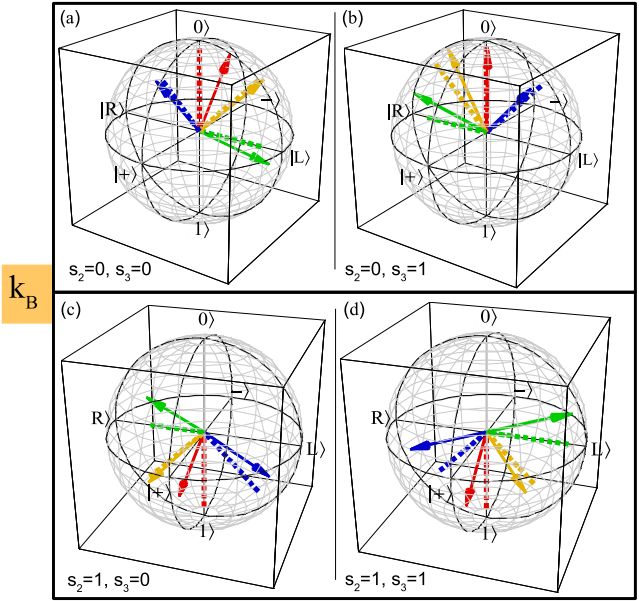}
	\end{center}
		\caption{(color online). Momentum ($\bf k_B$) output Bloch vectors of single qubit rotations. 
The experimental results (arrows) are shown with their projections on theoretical directions (dashed lines). 
Arrow colours correspond to different values of $\alpha$ and $\beta$ (see table \ref{table:rotation_k}). }
\label{fig:block_k}
\end{figure}
\begin{table}
	\begin{ruledtabular}
				\centering
		\begin{tabular}{cccc}
			&$\alpha (\beta=0)$ & F $(s_2=0,s_3=0)$ & F $(s_2=0,s_3=1)$ 
			\\
			\cline{1-4}					
			&$0$ & $0.961\pm0.003$ & $0.971\pm0.003$ 
			\\
$\bf k_B$&$\pi/2$ & $0.879\pm0.006$ & $0.895\pm0.005$ 
			\\
			&$\pi/4$ & $0.998\pm0.005$ & $0.961\pm0.006$ 
			\\	
			&$-\pi/4$ & $0.833\pm0.007$ & $0.956\pm0.006$ 
	\end{tabular}
	\begin{tabular}{cccc}
			&$\alpha (\beta=0)$ & F $(s_2=1,s_3=0)$ & F $(s_2=1,s_3=1)$ 
			\\
			\cline{1-4}					
			&$0$ & $0.944\pm0.0044$ & $0.943\pm0.005$ 
			\\
$\bf k_B$&$\pi/2$ & $0.799\pm0.007$ & $0.918\pm0.005$ 
			\\
			&$\pi/4$ & $0.919\pm0.008$ & $0.857\pm0.009$ 
			\\	
			&$-\pi/4$ & $0.946\pm0.008$ & $0.872\pm0.008$
	\end{tabular}
\end{ruledtabular}
		\caption{Momentum ($\bf k_B$) 
		experimental fidelities (F) of single qubit rotation output states for different values of $\alpha$ and $\beta$. 
		Each datum is obtained by the measurements of the different Stokes parameters, each one lasting 10 sec. 
}
\label{table:rotation_k}
\end{table}

We first realized the experiment without FF corrections (in this case we didn't use the retardation fiber and the Pockels cells in the setup).
The results obtained for $s_2=s_3=0$ (i.e. when the computation
proceeds without errors) with $\ket{\chi_{in}}=\ket{+}$ are shown in fig. \ref{fig:bloch_pi}(a). 
We show on the Bloch sphere the experimental output qubits and their projections on the theoretical state 
$HR_x(\beta)R_z(\alpha)\ket{+}$ for the whole set of $\alpha$ and $\beta$. 
The corresponding fidelities are given in table \ref{table:rotation_pi}. 
We also performed the tomographic analysis (shown in fig \ref{fig:bloch_pi}b),c),d)) on the output qubit $\pi_B$ for all the possible combinations of $s_2$ and $s_3$ 
and for the input qubit $\ket{\chi_{in}}=\ket{\pm}$. 
The high fidelities obtained in these measurements indicate that 
deterministic QC can be efficiently implemented in this configuration by 
Pauli errir active FF corrections.

They were realized by using the entire
measurement apparatus of fig. \ref{fig:schema}. Here two fast
driven transverse $LiNbO_{3}$ Pockels cells ($\sigma_x$ and $\sigma_z$)
with risetime $=1n\sec $ and $V_{\frac{\lambda }{2}}\thicksim 1KV$ 
are activated by the output signals of detectors $a_i$ ($i=1,3,4$) corresponding to the different
values of $s_{\pi_A}$ and $s_{\kk_A}$.
They perform the operation ${\sigma _{z}}^{s_{\pi_A}}{\sigma _{x}}^{s_{\kk_A}}$ on photon B, coming
from the output $s_{\kk_B}=0$ of $BS$ and transmitted through a
single mode optical fiber. Note that no correction is needed when
photon A is detected on the output $a_2$ ($s_{\pi_A}=s_{\kk_A}=0$). 
Temporal synchronization between the activation of the high voltage 
signal and the transmission of photon B through the Pockels cells is guaranteed 
by suitable choice of the delays $D$. We used only one $BS$ output of photon $B$, namely $s_{\kk_B}=0$,
in order to perform the algorithm with initial state $\ket{\chi_{in}}=\ket+$. 
The other $BS$ output corresponds to the algorithm starting with the initial state
$\ket{\chi_{in}}=\ket-$.
By referring to Fig. \ref{fig:schema}, each detector $a_j$ corresponds to a different value of $s_{\kk_A}$ and $s_{\pi_A}$.
Precisely, $a_1$ corresponds to $s_{\kk_A}=0$ and $s_{\pi_A}=1$ and activates the Pockels cell $\sigma_z$ (see eq. \ref{out}).
Detector $a_2$ corresponds to $s_{\kk_A}=s_{\pi_A}=0$, i.e. the computation without errors and thus no Pockels cell is activated.
Detector $a_3$ corresponds to $s_{\kk_A}=1$, $s_{\pi_A}=0$ and activates $\sigma_x$, while
$a_4$ corresponds to $s_{\kk_A}=s_{\pi_A}=1$ and both $\sigma_x$ and $\sigma_z$ are activated.
\begin{figure}[t]
\begin{center}
\includegraphics[width=8.5cm]{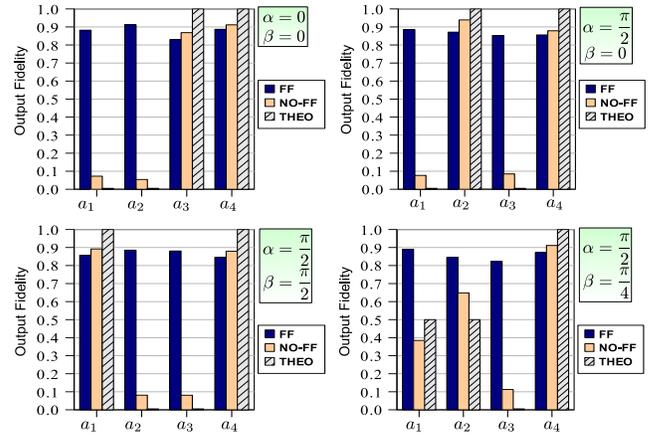}
\end{center}
\caption{(color online). Output fidelities of the single qubit rotation algorithm with [FF, black (blue) columns] or without
[NO FF, gray (orange) columns] feed forward. In both cases, the four columns of the histograms refer to the measurement of the
output state (encoded in the polarization of photon B) by detector $b_1$ in
coincidence with $a_1,\dots,a_4$ respectively. Grey dashed columns (THEO) correspond to theoretical fidelities in the no-FF case.
}
\label{fig:risultati}
\end{figure}

In fig. \ref{fig:risultati} the output state fidelities obtained 
with/without active FF corrections
(i.e. turning on/off the Pockels cells) 
are compared for different values of $\alpha$ and $\beta$. 
The expected theoretical fidelities in the no-FF case are also shown. 
In all the cases the computational errors are corrected by the FF action,
with average measured fidelity $F=0.867\pm0.018$. The overall repetition rate is about $500Hz$, 
which is more than 2 orders of magnitude larger than 
one-way single qubit-rotation realized 
with 4-photon cluster states.

We also demonstrated the computational equivalence of the two DOF's of photon $B$ by performing the same algorithm with the following qubit ordering:
\eq\label{order_b}
\begin{aligned}
b)\qquad \text{(1,2,3,4)=}&(\pi_B,\pi_A,{\bf k}_A,{\bf k}_B), \\
\mathcal U=&H\otimes\sigma_z\otimes\sigma_x\otimes \sigma_zH\,.
\end{aligned}
\fine
In this case we used the momentum of photon $B$ ($\kk_B$) as output state.
The explicit expression of the output state $|\chi_{out}\rangle_{\mathbf{k}_B}$ 
in the laboratory basis is now
$|\chi_{out}\rangle_{\mathbf{k}_B}={(\sigma_z)}^{s_3}{(\sigma_x)}
^{s_2}\sigma_zHR_x(\beta)R_z(\alpha)|\chi_{in}\rangle\,$.
By using only detectors $a_2$, $a_3$, $b_1$, $b_2$ in fig. \ref{fig:schema} we measured $|\chi_{out}\rangle_{\mathbf{k}_B}$ 
for different values of $\alpha$ (which correspond in
the laboratory to the polarization measurement bases $|\alpha_{\mp}\rangle_{\pi_A}$) and 
$\beta=0$ (which correspond in the laboratory to the momentum bases $|-\beta_{\pm}\rangle_{\mathbf{k}_A}$). 
The first qubit ($\pi_B$) was
always measured in the basis $|\pm\rangle_{\pi_B}$. The $\mathbf{k}_B$ tomographic analysis for all the possible 
values of $s_2\equiv s_{\pi_A}$ and $s_3\equiv s_{\kk_A}$ are shown in Fig. \ref{fig:block_k}.
{i.e. for different
values of $s_1\equiv s_{\pi_B}$)}. We obtained
an average value of fidelity $F>0.9$ (see table \ref{table:rotation_k}).
In this case the realization of deterministic QC by FF corrections could be realized by the adoption of active phase modulators.
The ($\pi $)-(${\mathbf{k}}$) computational equivalence and the use of active 
feed-forward show that the multidegree of freedom approach is feasible for deterministic one-way QC.

\subsection{{\sc c-not} gate}
\begin{figure}[t]
	\begin{center}
		\includegraphics[width=8cm]{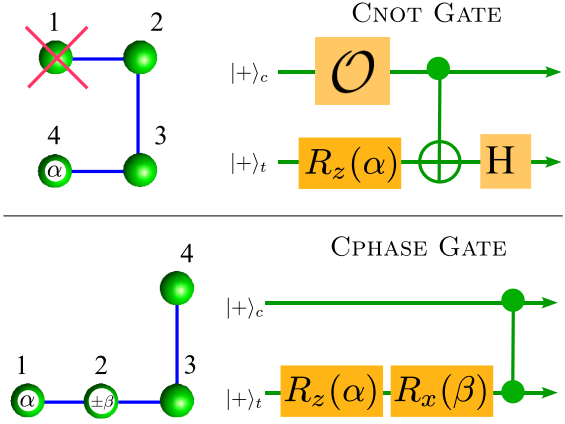}
	\end{center}
		\caption{(color online). {\sc c-not} and {\sc c-phase} gates by the four qubit cluster. (top) {\sc c-not} gate realization 
		via measurement of qubits 1, 4 on the horseshoe cluster and equivalent circuit.
		(bottom) {\sc c-phase} gate realization via measurement of qubits $1$ and $2$ in the bases $B_1(\alpha)$ and $B_2(\alpha)$ 
		and equivalent circuit.}
\label{fig:gates}
\end{figure}
The four qubit cluster allows the implementation of nontrivial two-qubit operations, such as the \textsc{c-not} gate.
Precisely, a {\sc c-not} gate acting on a generic target qubit belonging to the equatorial plane of the Bloch sphere 
(i.e. a generic state of the form $\frac1{\sqrt2}(\ket0+e^{i\gamma}\ket1)$) can be realized 
by the four-qubit horseshoe (180$^\circ$ rotated) cluster state $\cluster$ (see eq. \eqref{4qubit-cluster})
Let's consider Fig. \ref{fig:gates}(top).
The measurement of qubits 1 and 4 realizes a {\sc c-not} gate (the logical circuit shown in figure)
between the control ${\ket{+}}_c$ and target ${\ket{+}}_t$ qubit. 
By measuring the qubit 1 in the basis $\{{\ket0}_1,{\ket1}_1\}$ or ${\ket{\pm}}_1$
we realized on the control input qubit ${\ket{+}}_c$ the gate 
$\mathcal O=\openone$ or $\mathcal O=H$ respectively.
The measurement of qubit 4 in the basis $\ket{\alpha_{\pm}}_4$ realizes the gate $HR_z(\alpha)$ on the target input qubit ${\ket+}_t$.
The algorithm is concluded by the vertical link that perform a $CZ$ operation.
The input state $\ket{+}_c\otimes\ket{+}_t$ is transformed, in case of no ``errors'' (i.e. $s_1=s_4=0$),
into $\ket{\Psi_{out}}=CZ_{ct}(\mathcal O\ket{+}_c\otimes H_tR_z(\alpha)\ket{+}_t)=
H_t\text{\sc c-not}(\mathcal O\ket{+}_c\otimes R_z(\alpha)\ket{+}_t)$.
This circuit realizes the {\sc c-not} gate (up to the Hadamard $H_t$)
for arbitrary equatorial target qubit (since $R_z(\alpha)\ket+=\frac{e^{-i\alpha/2}}{\sqrt2}(\ket0+e^{i\alpha}\ket1)$)
and control qubit $\ket0,\ket1$ or $\ket{\pm}$ depending on the
measurement basis of qubit 1.
\begin{table}[t]
\begin{ruledtabular}
\centering
\begin{tabular}{c|c|c|c|c}
$\quad\mathcal O\quad$ & $\alpha$ & $\text{Control output}$ &  $F(s_4=0)$ & $F(s_4=1)$
  \\
  \hline\hline
  &   $\pi/2$ & $s_1=0\rightarrow\ket1_c$  & $0.965\pm0.004$ & $0.975\pm0.004$
  \\
  \cline{3-5}
  $H$          &    &$s_1=1\rightarrow\ket0_c$  & $0.972\pm0.004$ & $0.973\pm0.004$
  \\
  \cline{2-5}
  & $\pi/4$  &$s_1=0\rightarrow\ket1_c$  & $0.995\pm0.008$ & $0.902\pm0.012$
			\\
			\cline{3-5} 
      &			&$s_1=1\rightarrow\ket0_c$   & $0.946\pm0.010$ & $0.945\pm0.009$
			\end{tabular}
			\begin{tabular}{c|c|c|c|c}
  		$\ \mathcal O\ $ & $\alpha$ & $\text{Control output}$ &  $F(s_1=s_4=0)$ & $F(s_1=0,s_4=1)$
	 		\\
	 		\hline\hline
  		&  $\pi/2$  &$\ket0_c\equiv\ket \el_{\bf k_B}$  & $0.932\pm0.004$ & $0.959\pm0.003$
			\\
			\cline{3-5}
			$\openone$&&$\ket1_c=\ket r_{\bf k_B}$    & $0.941\pm0.005$ & $0.940\pm0.005$
			\\
			\cline{2-5} 
			&  $\pi/4$  &$\ket0_c=\ket \el_{\bf k_B}$  			& $0.919\pm0.007$ & $0.932\pm0.007$
			\\
			\cline{3-5}
			& &$\ket1_c=\ket r_{\bf k_B}$    	& $0.878\pm0.009$ & $0.959\pm0.006$
		\end{tabular}
		\end{ruledtabular}
		\caption{Experimental fidelity (F) of {\sc c-not} gate output target qubit for different value of $\alpha$ and $\mathcal O$.} 
		\label{table:CNOT}
\end{table}

The experimental realization of this gate is performed by adopting the following ordering between the physical qubits:
\begin{equation}
\begin{aligned} 
c)\qquad \text{(1,2,3,4)=}&({{\kk}}_A,{{\kk}}_B,\pi_B,\pi_A), \\
\mathcal U=&\sigma_zH\otimes\sigma_x\otimes\openone\otimes H\,.
\end{aligned}
\end{equation}
In this case the control output qubit is encoded in the momentum $\bf k_B$, while the target
output is encoded in the polarization $\pi_B$. 
In order to compensate the $H_t$ gate arising from the cluster algorithm 
we inserted two Hadamard in the polarization analysis of the detectors $b_1$ and $b_2$ (see fig. \ref{fig:schema}).
The output state in the laboratory basis is then written as 
\eq\label{cnot_out}
\ket{\Psi_{out}}=(\Sigma)^{s_4}\sigma^{(c)}_x
\text{\sc c-not}(\mathcal O\sigma^{s_1}_z\ket{+}_c\otimes R_z(\alpha)\ket{+}_t)\,,
\fine
where all the possible measurement outcomes of qubits 1 and 4 are considered. The Pauli errors are
$\Sigma=\sigma^{(c)}_z\sigma^{(t)}_z$, while the matrix $\sigma^{(c)}_x$ is due to the changing
between cluster and laboratory basis.

By measuring $\bf k_A$ in the basis $\{{\ket\ell}_{\kk_A},{\ket r}_{\kk_A}\}$ we perform the $\mathcal O=H$ operation on the control qubit.
By looking at eq. \eqref{cnot_out}, this means that if $s_{\kk_A}\equiv s_1=0$ ($s_{\kk_A}=1$) the control qubit is $\ket{1}$ ($\ket{0}$),
while the target qubit is $R_z(\alpha)\ket{+}_t$ ($\sigma_xR_z(\alpha)\ket{+}_t$).
In this case the gate acts on a control qubit $\ket0$ or $\ket1$, without any superposition of these two states.
We first verified that the gate acts correctly in this situation. In Table \ref{table:CNOT}(top) we report the experimental
fidelities ($F$) of the output target qubit $\pi_B$ for two different values of $\alpha$ and for the two possible values of $s_4$.
The high values of $F$ show that the gate works correctly when the control qubit is $\ket0$ or $\ket1$.

The second step was to verify that the gate works correctly with the control qubit in a superposition of $\ket0$ and $\ket1$.
This was realized by measuring the qubit $\bf k_A$ in the basis ${\ket\pm}_{\kk_A}$ and performing the 
$\mathcal O=\openone$ operation on the control qubit. The output state is written (without errors) as
$\ket{\Psi_{out}}=\ket{1}_c\otimes R_z(\alpha)\ket{+}_t+\ket{0}_c\otimes \sigma_xR_z(\alpha)\ket{+}_t$.
In Table \ref{table:CNOT}(bottom) we show the values of the experimental fidelities of the target qubit $\pi_B$,
corresponding to the measurement of the output control qubit $\kk_B$ in the basis $\{\ket0,\ket1\}$ when $\mathcal O=\openone$.
The results demonstrate the high quality of the operation also in this case.

\subsection{{\sc c-phase} gate}
\begin{table}[t]
\begin{ruledtabular}
\centering
    \begin{tabular}{cccc}
    $\alpha$ & $\beta$ & $F_{\kk_B}$ ($\kk_A=\ket-$) & $F_{\kk_B}$ ($\kk_A=\ket+$)\\
    0 & 0 & $0.878\pm0.004$ & $0.933\pm0.003$\\
    $\pi$ & 0 & $0.919\pm0.003$ & $0.917\pm0.004$\\
    $\pi/2$ & 0 & $0.876\pm0.005$ & $0.816\pm0.005$\\
    $-\pi/2$ & 0 & $0.880\pm0.004$ & $0.883\pm0.004$\\
    $\pi/2$ & $\pi/2$ & $0.969\pm0.002$ & $0.949\pm0.003$\\
    $\pi/2$ & $-\pi/2$ & $0.950\pm0.003$ & $0.939\pm0.003$\\
    $\pi/4$ & $\pi/2$ & $0.885\pm0.006$ & $0.916\pm0.005$\\
		\end{tabular}
		\end{ruledtabular}
		\caption{Experimental fidelity (F) of {\sc c-phase} gate output target qubit for different value of $\alpha$ and $\beta$. 
    In parenteses we indicate the corresponding measured output of the control qubit $\kk_A$.} 
		\label{table:cphase}
\end{table}
The four qubit cluster allows also the realization of a {\sc c-phase} gate for arbitrary target qubit
and fixed control $|+\rangle_c$.
The measurement pattern needed for this gate is shown in Fig. \ref{fig:gates}(bottom)
and consists of the measurements of qubits 1 and 2 in the bases $B_1(\alpha)$ and $B_2[(-1)^{s_1}\beta]$.
These two measurements realize a generic rotation $R_x(\beta)R_z(\alpha)$ on the input target qubit ${\ket+}_t$,
as explained in subsection \ref{subsec:rotations}. The link existing between qubit $3$ and $4$ in the cluster performs
the subsequent {\sc c-phase} gate between the control qubit ${\ket+}_c$ and a generic target qubit $R_x(\beta)R_z(\alpha){\ket+}_t$,

The experimental realization is done by considering the following ordering between the physical qubits:
\begin{equation}
\begin{aligned} 
d)\qquad\text{(1,2,3,4)=}&(\pi_A,\pi_B,{{\kk}}_B,{{\kk}}_A), 
\\
\mathcal U=&H\otimes\openone\otimes\sigma_x\otimes\sigma_z H. 
\end{aligned}
\end{equation}
We realized a {\sc c-phase} gate for arbitrary target qubit
and fixed control $|+\rangle_c$ (see Fig. \ref{fig:gates}(c)) by measuring 
qubits 1 and 2 of $\cluster$ in the bases 
$|\alpha_\pm\rangle$ and $\ket{(-)^{s_1}\beta_\pm}$ respectively. 
By considering ordering d) we encoded the output state in
the physical qubits $\mathbf{k}_A$ and $\mathbf{k}_B$. For $s_1=s_2=0$, by
using the appropriate base changing, the output state is written as 
\begin{equation}
|\Psi_{out}\rangle=|-\rangle_{\mathbf{k}_A}\otimes\sigma_x|\Phi\rangle_{%
\mathbf{k}_B}+|+\rangle_{\mathbf{k}_A}\otimes\sigma_x\sigma_z|\Phi\rangle_{%
\mathbf{k}_B}\,.
\end{equation}
Here $\ket{\Phi}_{{\bf k}_B}=R_x(\beta)R_z(\alpha)\ket+$ and the matrix $\sigma_x$ is due to the basis changing.
The measured fidelity of the target ${\bf k}_B$ corresponding to a control $\ket+_{{\bf k}_A}$ ($\ket-_{{\bf k}_A}$)
for different values of $\alpha$ and $\beta$ are shown in table \ref{table:cphase}.
We obtaining an average value $F=0.907\pm0.010$ ($F=0.908\pm0.011$).

\section{Algorithms}\label{sec:algorithms}

\subsection{Grover algorithm}
The Grover's search algorithm for two input qubits is implemented by using the four qubit 
cluster state \cite{97-gro-qua,97-gro-qua2,05-wal-exp, 07-pre-hig}.
\begin{figure}
	\centering
	\includegraphics[width=8.5cm]{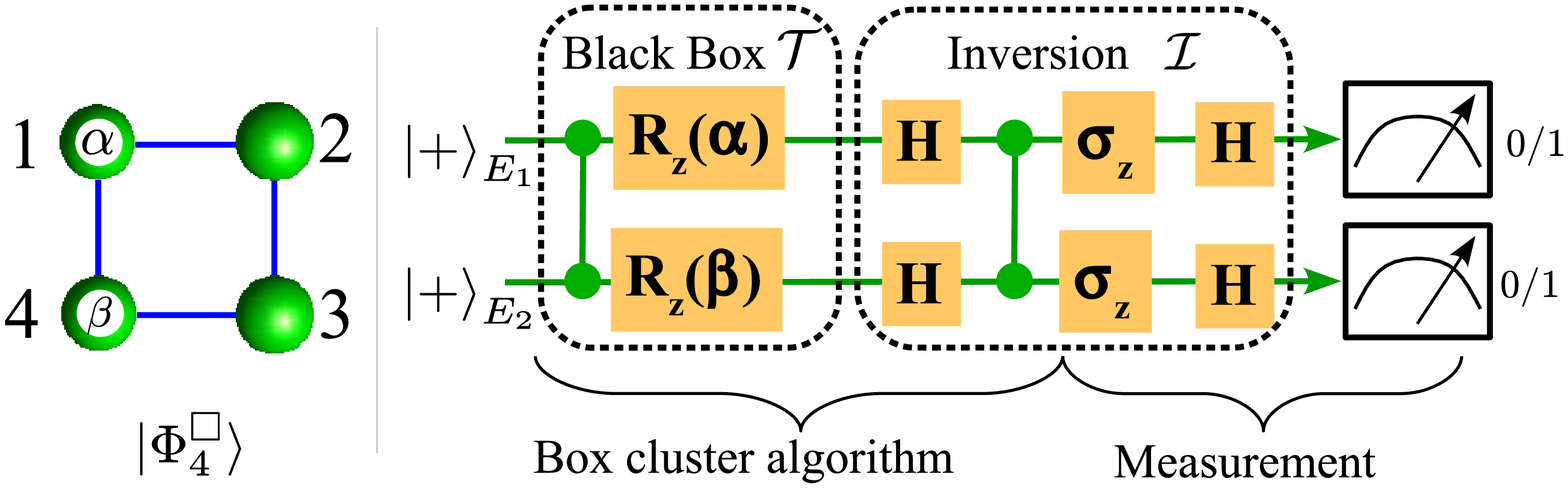} 
	\caption{(color online). Grover algorithm. Left: realization of the Grover algorithm through the measurements of qubit 1 and 4 on the box cluster $\boxcluster$.
        Right: logical circuit implementing the Grover operator $G=\mathcal I\cdot \mathcal T$. The first part of G is implemented by the cluster 
algorithm while the second part 
        by the measurement.}
	\label{fig:grover}
\end{figure}

Let's describe the algorithm in general. Suppose to have $2^M$ elements (encoded in $M$ qubits) and a black box (or oracle) that tags one of them.
The tagging, denoted as $\mathcal T$, is realized by changing the sign of the desired element.
The goal is to identify the tagged item by repeated query to the black box; the Grover's algorithm requires 
$\mathcal O(\sqrt{2^M})$ operations, while the best classical algorithm takes on average $2^M/2$ calculations.

The general algorithm starts with the input state prepared as $\ket{\Psi^+}\equiv{\ket+}_{E_1}\cdots{\ket+}_{E_M}$ 
and consists of repeated applications of the Grover operator $G$, given by the oracle tagging $\mathcal T$ followed
by the so called \textit{inversion about average} operation $\mathcal I\equiv2\ket{\Psi^+}\bra{\Psi^+}-\openone$. 
We can thus write $G\equiv\mathcal I\cdot \mathcal T$. In general after $R=\mathcal O(\sqrt{2^M})$
iterations of $G$ the tagged item is obtained at the output of the circuit with high probability.

In the case of $2$ qubits the quantum algorithm (shown in Fig. \ref{fig:grover}(right)) requires just one $G$ operation. 
The four elements are ${\ket0}_{E_1}{\ket0}_{E_2}$, ${\ket0}_{E_1}{\ket1}_{E_2}$, ${\ket1}_{E_1}{\ket0}_{E_2}$ and ${\ket1}_{E_1}{\ket1}_{E_2}$.
They are prepared in a complete superposition, i.e. in the state ${\ket+}_{E_1}{\ket+}_{E_2}$,
while the black box tagging acts simply by changing the sign to one of the elements, for instance $\ket1\ket0\rightarrow-\ket1\ket0$.
It consists of a {\sc c-phase} gate followed by two single qubit rotations, $R_z(\alpha)_1$ and $R_z(\beta)_2$.
By setting the rotation angles $\alpha\beta$ to $00,\pi0,0\pi$ or $\pi\pi$ the black box tags respectively the states
$\ket1\ket1$, $\ket1\ket0$, $\ket0,\ket1$,  or $\ket0\ket0$ (remember that $R_z(\pi)$ is $\sigma_z$ up to a global phase).
The inversion operation consists of a {\sc c-phase} gate and single qubit gates (see Fig. \ref{fig:grover}(right)).
The inversion acts such as the output state of the system is exactly the tagged item.

This algorithm can be realized in the one-way model by using the four-qubit box cluster previously defined. By measuring qubit $1$ and $4$ 
in the basis $B_1(\alpha)$ and $B_4(\beta)$ we implement the black box
and the first part of the inversion algorithm (\textit{Box cluster algorithm} in Fig. \ref{fig:grover}). 
The output qubits are then encoded into the physical qubit $2$ and $3$.
The $H\sigma_z$ operation needed to conclude the inversion operation can be performed at the measurement stage. Indeed we can measure the
qubit 2 and 3 in the basis $B_j(\pi)$: this is equivalent to apply $H\sigma_z$ gates and then perform the measurement in the computational basis
$\{\ket0,\ket1\}$ (see \textit{Measurement} in Fig. \ref{fig:grover}).

Without Pauli errors the desired tagged item is given by
$\ket{s_2}\ket{s_3}$.
Depending on the measurement outcome ($s_1$ and $s_4$) the corresponding Pauli errors are $(\sigma_z)^{s_1}(\sigma_x)^{s_4}$ on the
qubit $E_1$ and $(\sigma_z)^{s_4}(\sigma_x)^{s_1}$ on the qubit $E_2$. However, since the output of the algorithm will be
one of the four states of computational basis, the $\sigma_z$ operation leaves the output unchanged, while $\sigma_x$ flips the output state
(see equation\eqref{pauli_error}).
In this way the tagged item is found to be ${\ket{s_2\oplus s_4}}_{E_1}{\ket{s_3\oplus s_1}}_{E_2}$ and the FF corrections are simply
relabeling FF.
\begin{figure}
	\centering
	\includegraphics[width=8cm]{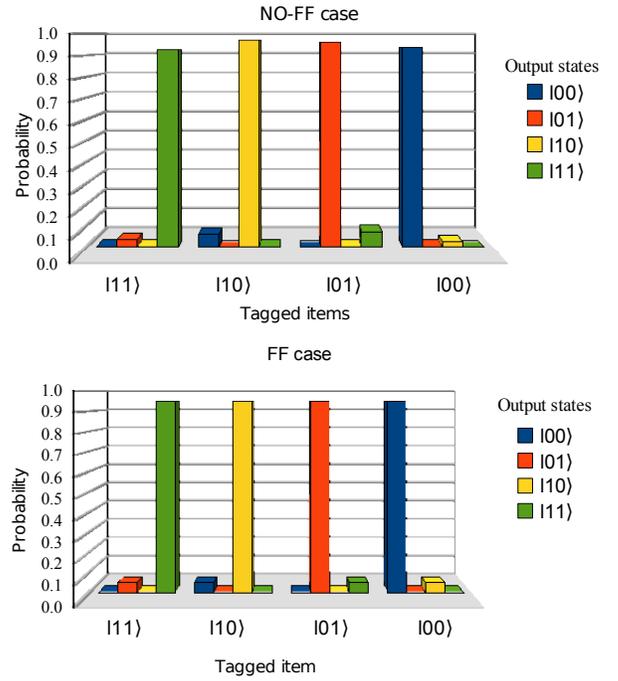} 
	\caption{(color online). Experimental results of the Grover algorithm. Upper graph: we report, for different tagged item, the probability of the different output states
        when the computation proceeds without Pauli errors. Experimental errors are of the order of $0.005$ for higher histograms, while for the lower ones becomes
        $0.0005$. Lower graph: experimental probabilities in the FF case. Experimental errors are of the order of $0.002$ for higher histograms, 
        while for the lower ones becomes      $0.0004$. 
        Each datum is obtained by $10s$ measurement. }
	\label{fig:grover_result}
\end{figure}

Let's now describe the experimental realization of the Grover algorithm by our apparatus. 
If we consider the following ordering of the physical qubits (see equation \eqref{equiv cluster}):
\eq
\begin{aligned}
e)\quad \text{(1,2,3,4)=}&({\bf k}_B,\pi_A,{\bf k}_A,\pi_B), \\
\mathcal U=&\sigma_xH\otimes H\otimes\sigma_zH\otimes H\,,
\end{aligned}
\fine
the generated state \eqref{cluster} is equivalent to the box cluster $\boxcluster$ up to the single qubit unitaries given by $\mathcal U$.
By this change of basis we can determine the correct measurement to be performed in the laboratory basis.

The experimental results are shown in Fig. \ref{fig:grover_result}. In the upper graph we show the experimental fidelities obtained
when the computation proceeds without Pauli errors, i.e. $s_1=s_4=0$. The mean probability to identify the tagged item
is $0.9482\pm0.0080$ and the algorithm is realized at $\sim250Hz$. We report in the lower graph the probability of identification
when the FF corrections are implemented. The tagged item is discovered with probability $0.9475\pm0.0022$ and the algorithm
is realized at $\sim 1kHz$ repetition rate, as expected. Note that, in the lower graph, a change of the tagged item corresponds
to reorder the histograms. This is due to the fact that the measurement in the basis $B(\pi)=\{\ket-,\ket+\}$ is
the same as $B(0)=\{\ket+,\ket-\}$: the difference is that in the first case we associate $s=0$ to $\ket-$ while
in the second we associate $s=0$ to $\ket+$.

\subsection{Deutsch algorithm}
\begin{figure}[t]
	\begin{center}
		\includegraphics[width=8.5cm]{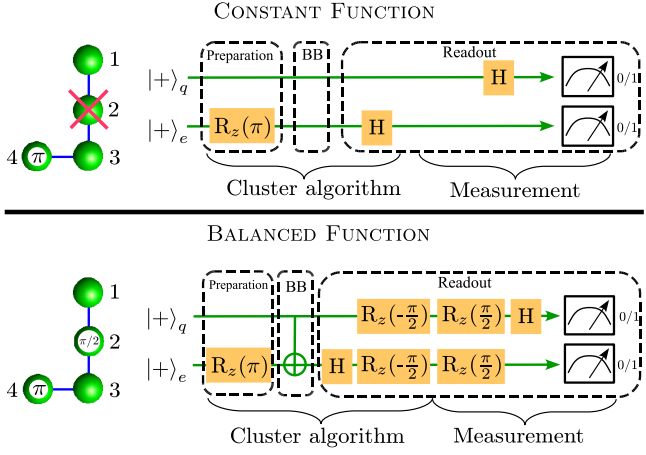}
	\end{center}
		\caption{(color online). Implementation of the Deutsch algorithm. Measurement pattern on the physical cluster (left) and corresponding
    operations on the logical qubits (right). Measurements of qubit $2$ and $4$ on the left correspond to the \textit{Cluster algorithm} in the right.
    The output logical qubits are encoded in the physical qubits $1$ (query, $q$) and $3$ (ancilla, $e$) 
    that should be measured in order to read out the answer of the algorithm 
    (\textit{Measurement} in the right).}
\label{fig:deutsch}
\end{figure}
\begin{table}[t]
  \begin{ruledtabular}
  \centering
    \begin{tabular}{ccccc}
      &  \multicolumn{2}{c}{Constant functions}& \multicolumn{2}{c}{Balanced functions}\\
      & $f_1$ & $f_2$ & $f_3$ & $f_4$\\
      BB & $\openone_a\otimes\openone_e$ & $\openone_a\otimes\sigma^{[e]}_x$ & CNOT$_{ae}$ & $(\openone_a\otimes\sigma^{[e]}_x)$CNOT$_{ae}$
    \end{tabular}
	\end{ruledtabular}
  \caption{Oracle operation (BB) depending on the single qubit function $f_i$. The two function $f_1$ and $f_2$ are constant, while
  $f_3$ and $f_4$ are balanced.} 
  \label{table:BB}
\end{table}
The four qubit cluster state allow the implementation of the Deutsch's algorithm for two input qubits
\cite{07-tam-exp}. This algorithm distinguishes two kinds of functions $f(x)$ acting on a generic $M$-bit query input:
the \textit{constant} function returns the same value ($0$ or $1$) for all input $x$ and the \textit{balanced} function gives
$0$ for half of the inputs and $1$ for other half. Usually the function is implemented by a black box (or oracle).
By the Deutsch's algorithm one needs to query the oracle just once, while by using deterministic classical algorithms one needs to
know the output of the oracle many times (as $2^{M-1}+1$).
The oracle implements the function $f$ on the query input ${\ket x}_q$ through an ancillary qubit ${\ket y}_e$:
\begin{figure}[t]
	\begin{center}
		\includegraphics[width=7cm]{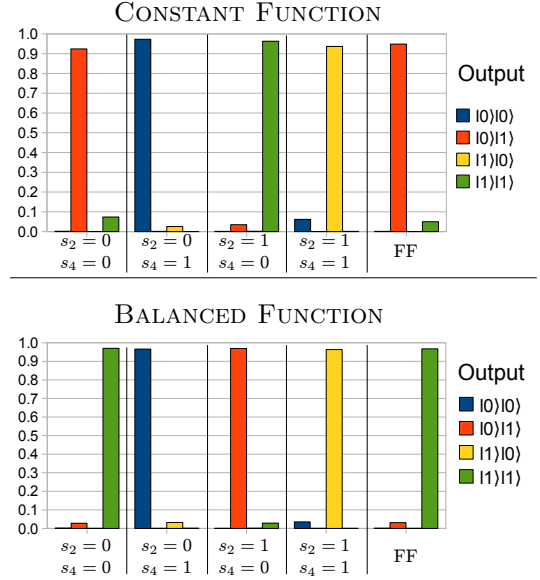}
	\end{center}
		\caption{(color online). Experimental probabilities of the output states of the Deutsch algorithm in case of constant (top) or balanced (bottom) function.
		The probabilities are shown for all the values of $s_4$ and $s_2$, while in the last column are shown the results after the FF relabeling operation.}
\label{fig:deutsch_result}
\end{figure}
\eq\label{oracle}
\text{Oracle:}{\ket{x}}_q{\ket{y}}_{e}\rightarrow{\ket{x}}_q{\ket{y\oplus f(x)}}_e\,,
\fine
where $y=0,1$ and $x=0,1,\cdots,2^M-1$. If the oracle acts on the input qubits ${\ket{+}}_1{\ket+}_2\cdots{\ket+}_M{\ket{-}}_e$, the
output state is
\eq
\frac1{\sqrt{2^{M}}}\sum^{2^M-1}_{x=0}(-1)^{f(x)}{\ket x}_q{\ket-}_e\,.
\fine
By applying Hadamard gates for each qubits the output state can be written as
\eq\label{output_deutsch}
\begin{cases}
\begin{aligned}
\bigotimes^M_{a=1}{\ket{0}}_a{\ket1}_e\qquad &\text{if }f\text{ is constant}
\\
\bigotimes^M_{a=1}{\ket{1}}_a{\ket1}_e\qquad &\text{if }f\text{ is balanced}
\end{aligned}
\end{cases}
\fine
Then by measuring the query state in the computational basis we can discover if the function $f$ is constant or balanced.
The algorithm thus proceeds in three steps: 
\begin{itemize}
\item \textit{preparation}: it consists in the initialization of the input state into ${\ket{+}}_1{\ket+}_2\cdots{\ket+}_M{\ket{-}}_e$.
\item \textit{BB}: this is the Oracle operation \eqref{oracle}.
\item \textit{readout}: apply Hadamard gates for each qubits and measure them in the computational basis $\{\ket{0},\ket{1}\}$.
\end{itemize}

In the two-qubit version the function $f$ acts on a single qubit $\ket{x}_{q}$.
In this case there are 4 possible functions $f$ on a single qubit: two are constant, namely $f_1(0)=f_1(1)=0$ and $f_2(0)=f_2(1)=1$,
while two are balanced $f_3(0)=0,f_3(1)=1$ and $f_4(0)=1,f_4(1)=0$.
Let's describe the oracle operation \eqref{oracle} as a ``Black Box'' (BB).
In Table \ref{table:BB} we give the oracle operation on the two qubits $\ket{x}_q\ket{y}_e$
depending on the chosen function $f_i$. 

By the four qubit cluster is possible to implement the two-qubit version of the algorithm.
Let's consider Fig. \ref{fig:deutsch}. The algorithm is implemented by the measurements of qubit $2$ and $4$, while the output 
is encoded in the qubit $1$ (query qubit) and $3$ (ancillary qubit). Precisely, the measurement of qubit $4$ in the basis $B_4(\pi)$ 
performs the transformation $HR_z(\pi)$ on the ancillary qubit.

The BB is implemented by the measurement of qubit $2$. If the oracle choose the measurement basis $\{\ket{0}_2,\ket{1}_2\}$
it implements the $f_1$ function. In fact the qubit $2$ is removed from the cluster and no operations are performed on the other qubits.
The full cluster algorithm consists in the operation $\openone_q\otimes[HR_z(\pi)]_e$.
This is exactly the Deutsch algorithm in the case of constant function up to an Hadamard gate on the query qubit to be implemented
in the final measurement step. The measurement basis on the output qubits $1$ and $3$ are then 
$B_1(0)$ and $\{\ket{0}_3,\ket1_3\}$ (see \textit{Measurement} in Fig. \ref{fig:deutsch}).

By choosing $B_2(\frac \pi2)$ as measurement basis for qubit $2$
the oracle implements the $f_3$ function and the obtained operation is (see CBB$_3$) 
$[R_z(-\frac\pi2)]_q\otimes[R_z(-\frac\pi2)]_e\cdot$CZ$_{qe}$. Together with the $\openone_q\otimes[HR_z(\pi)]_e$ the 
full cluster algorithm becomes
$[R_z(-\frac\pi2)]_q\otimes[R_z(-\frac\pi2)H]_e\cdot$CNOT$_{qe}\cdot\openone_q\otimes[R_z(\pi)]_e$ 
(the cluster algorithm in Fig. \ref{fig:deutsch}).
This is the Deutsch algorithm in case of balanced function up to $[R_z(-\frac\pi2)]_q\otimes[R_z(-\frac\pi2)H]_e$
to be corrected in the final measurement step. These corrections corresponds to the choice of the measurement basis on the output qubits $1$ and $3$ as
$B_1(-\frac\pi2)$ and $\{\ket{0}_3,\ket1_3\}$ (see \textit{Measurement} in Fig. \ref{fig:deutsch}).

Without Pauli errors the output state of the Deutsch algorithm is given by
$\ket{s_1=0}_q\ket{s_3=1}_e$ in case of $f_1$ and $\ket{s_1=1}_q\ket{s_3=1}_e$ in case of $f_3$ (see eq. \eqref{output_deutsch}).
The correct outcome considering the Pauli errors are in this case
\eq\label{deutsch_relabel}
\begin{aligned}
&{\ket{s_1\oplus s_2}}_{q}{\ket{s_3\oplus s_4}}_e\qquad &\text{for }f_1\\
&{\ket{s_1\oplus s_2\oplus s_4}}_{q}{\ket{s_3\oplus s_4}}_e\qquad&\text{for }f_3
\end{aligned}
\fine
Note that the BB operation obtained by the function $f_2$ ($f_4$) is essentially the same, up to a global phase,
with respect to the function $f_1$ ($f_3$). In the following we then show
only the results obtained in the case of $f_1$ and $f_3$.
 
In Fig. \ref{fig:deutsch_result} we show the experimental probabilities of the different output states in function of the 
value of $s_2$ and $s_4$ for the $f_1$ and $f_3$ case. 
In the case of $f_1$, for $s_2=s_4=0$ the output of the algorithm is the state $\ket0_q\ket1_e$ with probability $0.924\pm0.005$,
while in the other cases are $\ket{0\oplus s_2}_q\ket{1\oplus s_4}_e$ consistently with eq. \eqref{deutsch_relabel}.
By using the FF relabeling we obtain the correct output with probability $0.949\pm0.002$.
In the case of balanced function the correct output after the FF operation is obtained with probability $0.967\pm0.002$.

\section{Conclusions}\label{sec:conclusions}
We have described the basic principles of operation of a one-way quantum computer operating with cluster states that are formed by two photons 
entangled in two different DOF's. We have also presented the experiment (and the relative results) carried out when the DOF's 
are the polarization and the linear momentum.
One-way QC based on multi-DOF cluster states presents some important advantages with respect to the one performed with multiphoton cluster states. In particular:
\begin{itemize}
\item the repetition rate of computation is almost three order of magnitude larger; 
\item the fidelity of the computational operations is much higher (nearly 90\%, even with active FF);
\item in some cases the intermediate FF operations do not require, differently from the case of multiphoton cluster states, active modulators.
      The FF operations are automatically (and also deterministically) implemented in these cases because of the entanglement 
      existing between the two DOF's of the same particle;
\item working with two photons allows to minimize the problems caused by the limited value of the detection efficiency.
\end{itemize}

A larger number of qubits is necessary to perform more complex gates and algorithms. For instance, using a Type I NL crystal a continuum 
of $\kk$-emission modes is virtually available to create a multiqubit spatial entangled state. Even if the number of modes scales exponentially 
with the number of qubits, it is still possible to obtain a reasonable number (six or even eight) of qubits. 
Hence, increasing the number of DOF's of the photon allows to move further than what expected by increasing the number of photons.
Because of all these reasons we believe that cluster states based on many DOF's may represent a good solution for experimental QC with photons 
on a mid-term perspective.

{\begin{acknowledgments}
This work was supported by Finanziamento Ateneo 06.
\end{acknowledgments}}


\begin{thebibliography}{31}
\expandafter\ifx\csname natexlab\endcsname\relax\def\natexlab#1{#1}\fi
\expandafter\ifx\csname bibnamefont\endcsname\relax
  \def\bibnamefont#1{#1}\fi
\expandafter\ifx\csname bibfnamefont\endcsname\relax
  \def\bibfnamefont#1{#1}\fi
\expandafter\ifx\csname citenamefont\endcsname\relax
  \def\citenamefont#1{#1}\fi
\expandafter\ifx\csname url\endcsname\relax
  \def\url#1{\texttt{#1}}\fi
\expandafter\ifx\csname urlprefix\endcsname\relax\def\urlprefix{URL }\fi
\providecommand{\bibinfo}[2]{#2}
\providecommand{\eprint}[2][]{\url{#2}}

\bibitem[{\citenamefont{Raussendorf and Briegel}(2001)}]{01-rau-aon}
\bibinfo{author}{\bibfnamefont{R.}~\bibnamefont{Raussendorf}} \bibnamefont{and}
  \bibinfo{author}{\bibfnamefont{H.~J.} \bibnamefont{Briegel}},
  \bibinfo{journal}{Phys. Rev. Lett.} \textbf{\bibinfo{volume}{86}},
  \bibinfo{pages}{5188} (\bibinfo{year}{2001}).

\bibitem[{\citenamefont{Nielsen}(2004)}]{04-nie-opt}
\bibinfo{author}{\bibfnamefont{M.~A.} \bibnamefont{Nielsen}},
  \bibinfo{journal}{Phys. Rev. Lett.} \textbf{\bibinfo{volume}{93}},
  \bibinfo{pages}{040503} (\bibinfo{year}{2004}).

\bibitem[{\citenamefont{Browne and Rudolph}(2005)}]{05-bro-res}
\bibinfo{author}{\bibfnamefont{D.~E.} \bibnamefont{Browne}} \bibnamefont{and}
  \bibinfo{author}{\bibfnamefont{T.}~\bibnamefont{Rudolph}},
  \bibinfo{journal}{Phys. Rev. Lett.} \textbf{\bibinfo{volume}{95}},
  \bibinfo{pages}{010501} (\bibinfo{year}{2005}).

\bibitem[{\citenamefont{Kiesel et~al.}(2005)\citenamefont{Kiesel, Schmid,
  Weber, T\'oth, {G{\"u}hne}, Ursin, and Weinfurter}}]{05-kie-exp}
\bibinfo{author}{\bibfnamefont{N.}~\bibnamefont{Kiesel}},
  \bibinfo{author}{\bibfnamefont{C.}~\bibnamefont{Schmid}},
  \bibinfo{author}{\bibfnamefont{U.}~\bibnamefont{Weber}},
  \bibinfo{author}{\bibfnamefont{G.}~\bibnamefont{T\'oth}},
  \bibinfo{author}{\bibfnamefont{O.}~\bibnamefont{{G{\"u}hne}}},
  \bibinfo{author}{\bibfnamefont{R.}~\bibnamefont{Ursin}}, \bibnamefont{and}
  \bibinfo{author}{\bibfnamefont{H.}~\bibnamefont{Weinfurter}},
  \bibinfo{journal}{Phys. Rev. Lett.} \textbf{\bibinfo{volume}{95}},
  \bibinfo{pages}{210502} (\bibinfo{year}{2005}).

\bibitem[{\citenamefont{Walther et~al.}(2005)\citenamefont{Walther, Resch,
  Rudolph, Schenck, Weinfurter, Vedral, Aspelmeyer, and
  Zeilinger}}]{05-wal-exp}
\bibinfo{author}{\bibfnamefont{P.}~\bibnamefont{Walther}},
  \bibinfo{author}{\bibfnamefont{K.~J.} \bibnamefont{Resch}},
  \bibinfo{author}{\bibfnamefont{T.}~\bibnamefont{Rudolph}},
  \bibinfo{author}{\bibfnamefont{E.}~\bibnamefont{Schenck}},
  \bibinfo{author}{\bibfnamefont{H.}~\bibnamefont{Weinfurter}},
  \bibinfo{author}{\bibfnamefont{V.}~\bibnamefont{Vedral}},
  \bibinfo{author}{\bibfnamefont{M.}~\bibnamefont{Aspelmeyer}},
  \bibnamefont{and}
  \bibinfo{author}{\bibfnamefont{A.}~\bibnamefont{Zeilinger}},
  \bibinfo{journal}{Nature} \textbf{\bibinfo{volume}{434}},
  \bibinfo{pages}{169} (\bibinfo{year}{2005}).

\bibitem[{\citenamefont{Bodiya and Duan}(2006)}]{06-bod-sca}
\bibinfo{author}{\bibfnamefont{T.~P.} \bibnamefont{Bodiya}} \bibnamefont{and}
  \bibinfo{author}{\bibfnamefont{L.-M.} \bibnamefont{Duan}},
  \bibinfo{journal}{Phys. Rev. Lett.} \textbf{\bibinfo{volume}{97}},
  \bibinfo{pages}{143601} (\bibinfo{year}{2006}).

\bibitem[{\citenamefont{Danos and Kashefi}(2006)}]{06-dan-det}
\bibinfo{author}{\bibfnamefont{V.}~\bibnamefont{Danos}} \bibnamefont{and}
  \bibinfo{author}{\bibfnamefont{E.}~\bibnamefont{Kashefi}},
  \bibinfo{journal}{Phys. Rev. A} \textbf{\bibinfo{volume}{74}},
  \bibinfo{pages}{052310} (\bibinfo{year}{2006}).

\bibitem[{\citenamefont{Prevedel et~al.}(2007)\citenamefont{Prevedel, Walther,
  Tiefenbacher, B{\"o}hi, Kaltenbaek, Jennewein, and Zeilinger}}]{07-pre-hig}
\bibinfo{author}{\bibfnamefont{R.}~\bibnamefont{Prevedel}},
  \bibinfo{author}{\bibfnamefont{P.}~\bibnamefont{Walther}},
  \bibinfo{author}{\bibfnamefont{F.}~\bibnamefont{Tiefenbacher}},
  \bibinfo{author}{\bibfnamefont{P.}~\bibnamefont{B{\"o}hi}},
  \bibinfo{author}{\bibfnamefont{R.}~\bibnamefont{Kaltenbaek}},
  \bibinfo{author}{\bibfnamefont{T.}~\bibnamefont{Jennewein}},
  \bibnamefont{and}
  \bibinfo{author}{\bibfnamefont{A.}~\bibnamefont{Zeilinger}},
  \bibinfo{journal}{Nature} \textbf{\bibinfo{volume}{445}}, \bibinfo{pages}{65}
  (\bibinfo{year}{2007}).

\bibitem[{\citenamefont{Popescu}(2007)}]{07-pop-kni}
\bibinfo{author}{\bibfnamefont{S.}~\bibnamefont{Popescu}},
  \bibinfo{journal}{Phys. Rev. Lett.} \textbf{\bibinfo{volume}{99}},
  \bibinfo{pages}{250501} (\bibinfo{year}{2007}).

\bibitem[{\citenamefont{Varnava et~al.}(2008)\citenamefont{Varnava, Browne, and
  Rudolph}}]{08-var-how}
\bibinfo{author}{\bibfnamefont{M.}~\bibnamefont{Varnava}},
  \bibinfo{author}{\bibfnamefont{D.~E.} \bibnamefont{Browne}},
  \bibnamefont{and} \bibinfo{author}{\bibfnamefont{T.}~\bibnamefont{Rudolph}},
  \bibinfo{journal}{Phys. Rev. Lett.} \textbf{\bibinfo{volume}{100}},
  \bibinfo{pages}{060502} (\bibinfo{year}{2008}).

\bibitem[{\citenamefont{Lu et~al.}(2007)\citenamefont{Lu, Zhou, G{\"u}hne, Gao,
  Zhang, Yuan, Goebel, Yang, and Pan}}]{07-lu-exp}
\bibinfo{author}{\bibfnamefont{C.-Y.} \bibnamefont{Lu}},
  \bibinfo{author}{\bibfnamefont{X.-Q.} \bibnamefont{Zhou}},
  \bibinfo{author}{\bibfnamefont{O.}~\bibnamefont{G{\"u}hne}},
  \bibinfo{author}{\bibfnamefont{W.-B.} \bibnamefont{Gao}},
  \bibinfo{author}{\bibfnamefont{J.}~\bibnamefont{Zhang}},
  \bibinfo{author}{\bibfnamefont{Z.-S.} \bibnamefont{Yuan}},
  \bibinfo{author}{\bibfnamefont{A.}~\bibnamefont{Goebel}},
  \bibinfo{author}{\bibfnamefont{T.}~\bibnamefont{Yang}}, \bibnamefont{and}
  \bibinfo{author}{\bibfnamefont{J.-W.} \bibnamefont{Pan}},
  \bibinfo{journal}{Nature Phys.} \textbf{\bibinfo{volume}{3}},
  \bibinfo{pages}{91} (\bibinfo{year}{2007}).

\bibitem[{\citenamefont{Cabello}(2005)}]{05-cab-str}
\bibinfo{author}{\bibfnamefont{A.}~\bibnamefont{Cabello}},
  \bibinfo{journal}{Phys. Rev. Lett.} \textbf{\bibinfo{volume}{95}},
  \bibinfo{pages}{210401} (\bibinfo{year}{2005}).

\bibitem[{\citenamefont{Scarani et~al.}(2005)\citenamefont{Scarani, Acin,
  Schenck, and Aspelmeyer}}]{05-sca-non}
\bibinfo{author}{\bibfnamefont{V.}~\bibnamefont{Scarani}},
  \bibinfo{author}{\bibfnamefont{A.}~\bibnamefont{Acin}},
  \bibinfo{author}{\bibfnamefont{E.}~\bibnamefont{Schenck}}, \bibnamefont{and}
  \bibinfo{author}{\bibfnamefont{M.}~\bibnamefont{Aspelmeyer}},
  \bibinfo{journal}{Phys. Rev. A} \textbf{\bibinfo{volume}{71}},
  \bibinfo{pages}{042325} (\bibinfo{year}{2005}).

\bibitem[{\citenamefont{Vallone et~al.}(2007)\citenamefont{Vallone, Pomarico,
  Mataloni, {De Martini}, and Berardi}}]{07-val-rea}
\bibinfo{author}{\bibfnamefont{G.}~\bibnamefont{Vallone}},
  \bibinfo{author}{\bibfnamefont{E.}~\bibnamefont{Pomarico}},
  \bibinfo{author}{\bibfnamefont{P.}~\bibnamefont{Mataloni}},
  \bibinfo{author}{\bibfnamefont{F.}~\bibnamefont{{De Martini}}},
  \bibnamefont{and} \bibinfo{author}{\bibfnamefont{V.}~\bibnamefont{Berardi}},
  \bibinfo{journal}{Phys. Rev. Lett.} \textbf{\bibinfo{volume}{98}},
  \bibinfo{pages}{180502} (\bibinfo{year}{2007}).

\bibitem[{\citenamefont{Nielsen and Chuang}(2000)}]{00-nie-qua}
\bibinfo{author}{\bibfnamefont{M.}~\bibnamefont{Nielsen}} \bibnamefont{and}
  \bibinfo{author}{\bibfnamefont{I.}~\bibnamefont{Chuang}},
  \emph{\bibinfo{title}{{Quantum Computation and Quantum Information}}}
  (\bibinfo{publisher}{{Cambridge University Press}}, \bibinfo{year}{2000}).

\bibitem[{\citenamefont{Knill et~al.}(2001)\citenamefont{Knill, Laflamme, and
  Milburn}}]{01-kni-asc}
\bibinfo{author}{\bibfnamefont{E.}~\bibnamefont{Knill}},
  \bibinfo{author}{\bibfnamefont{R.}~\bibnamefont{Laflamme}}, \bibnamefont{and}
  \bibinfo{author}{\bibfnamefont{G.~J.} \bibnamefont{Milburn}},
  \bibinfo{journal}{Nature} \textbf{\bibinfo{volume}{409}}, \bibinfo{pages}{46}
  (\bibinfo{year}{2001}).

\bibitem[{\citenamefont{Kok et~al.}(2007)\citenamefont{Kok, Munro, Nemoto,
  Ralph, Dowling, and Milburn}}]{07-kok-lin}
\bibinfo{author}{\bibfnamefont{P.}~\bibnamefont{Kok}},
  \bibinfo{author}{\bibfnamefont{W.~J.} \bibnamefont{Munro}},
  \bibinfo{author}{\bibfnamefont{K.}~\bibnamefont{Nemoto}},
  \bibinfo{author}{\bibfnamefont{T.~C.} \bibnamefont{Ralph}},
  \bibinfo{author}{\bibfnamefont{J.~P.} \bibnamefont{Dowling}},
  \bibnamefont{and} \bibinfo{author}{\bibfnamefont{G.~J.}
  \bibnamefont{Milburn}}, \bibinfo{journal}{Rev. Mod. Phys.}
  \textbf{\bibinfo{volume}{79}}, \bibinfo{pages}{135} (\bibinfo{year}{2007}).

\bibitem[{\citenamefont{Barbieri et~al.}(2005)\citenamefont{Barbieri, Cinelli,
  Mataloni, and {De Martini}}}]{05-bar-pol}
\bibinfo{author}{\bibfnamefont{M.}~\bibnamefont{Barbieri}},
  \bibinfo{author}{\bibfnamefont{C.}~\bibnamefont{Cinelli}},
  \bibinfo{author}{\bibfnamefont{P.}~\bibnamefont{Mataloni}}, \bibnamefont{and}
  \bibinfo{author}{\bibfnamefont{F.}~\bibnamefont{{De Martini}}},
  \bibinfo{journal}{Phys. Rev. A} \textbf{\bibinfo{volume}{72}},
  \bibinfo{pages}{052110} (\bibinfo{year}{2005}).

\bibitem[{\citenamefont{Barbieri et~al.}(2006)\citenamefont{Barbieri, {De
  Martini}, Mataloni, Vallone, and Cabello}}]{06-bar-enh}
\bibinfo{author}{\bibfnamefont{M.}~\bibnamefont{Barbieri}},
  \bibinfo{author}{\bibfnamefont{F.}~\bibnamefont{{De Martini}}},
  \bibinfo{author}{\bibfnamefont{P.}~\bibnamefont{Mataloni}},
  \bibinfo{author}{\bibfnamefont{G.}~\bibnamefont{Vallone}}, \bibnamefont{and}
  \bibinfo{author}{\bibfnamefont{A.}~\bibnamefont{Cabello}},
  \bibinfo{journal}{Phys. Rev. Lett.} \textbf{\bibinfo{volume}{97}},
  \bibinfo{pages}{140407} (\bibinfo{year}{2006}).

\bibitem[{\citenamefont{Vallone
  et~al.}(2008{\natexlab{a}})\citenamefont{Vallone, Pomarico, {De Martini}, and
  Mataloni}}]{08-val-act}
\bibinfo{author}{\bibfnamefont{G.}~\bibnamefont{Vallone}},
  \bibinfo{author}{\bibfnamefont{E.}~\bibnamefont{Pomarico}},
  \bibinfo{author}{\bibfnamefont{F.}~\bibnamefont{{De Martini}}},
  \bibnamefont{and} \bibinfo{author}{\bibfnamefont{P.}~\bibnamefont{Mataloni}},
  \bibinfo{journal}{Phys. Rev. Lett.} \textbf{\bibinfo{volume}{100}},
  \bibinfo{pages}{160502} (\bibinfo{year}{2008}{\natexlab{a}}).

\bibitem[{\citenamefont{Vallone
  et~al.}(2008{\natexlab{b}})\citenamefont{Vallone, Pomarico, {De Martini}, and
  Mataloni}}]{08-val-one}
\bibinfo{author}{\bibfnamefont{G.}~\bibnamefont{Vallone}},
  \bibinfo{author}{\bibfnamefont{E.}~\bibnamefont{Pomarico}},
  \bibinfo{author}{\bibfnamefont{F.}~\bibnamefont{{De Martini}}},
  \bibnamefont{and} \bibinfo{author}{\bibfnamefont{P.}~\bibnamefont{Mataloni}},
  \bibinfo{journal}{Las. Phys. Lett.} \textbf{\bibinfo{volume}{5}},
  \bibinfo{pages}{398} (\bibinfo{year}{2008}{\natexlab{b}}).

\bibitem[{\citenamefont{Chen et~al.}(2007)\citenamefont{Chen, Li, Zhang, Chen,
  Goebel, Chen, Mair, and Pan}}]{07-che-exp}
\bibinfo{author}{\bibfnamefont{K.}~\bibnamefont{Chen}},
  \bibinfo{author}{\bibfnamefont{C.-M.} \bibnamefont{Li}},
  \bibinfo{author}{\bibfnamefont{Q.}~\bibnamefont{Zhang}},
  \bibinfo{author}{\bibfnamefont{Y.-A.} \bibnamefont{Chen}},
  \bibinfo{author}{\bibfnamefont{A.}~\bibnamefont{Goebel}},
  \bibinfo{author}{\bibfnamefont{S.}~\bibnamefont{Chen}},
  \bibinfo{author}{\bibfnamefont{A.}~\bibnamefont{Mair}}, \bibnamefont{and}
  \bibinfo{author}{\bibfnamefont{J.-W.} \bibnamefont{Pan}},
  \bibinfo{journal}{Phys. Rev. Lett.} \textbf{\bibinfo{volume}{99}},
  \bibinfo{pages}{120503} (\bibinfo{year}{2007}).

\bibitem[{\citenamefont{Nielsen}(2005)}]{05-nie-clu}
\bibinfo{author}{\bibfnamefont{M.~A.} \bibnamefont{Nielsen}},
  \emph{\bibinfo{title}{Cluster-state quantum computation}}
  (\bibinfo{year}{2005}),
  \eprint{\href{http://arxiv.org/pdf/quant-ph/0504097v2}{[quant-ph/0504097v2]}%
}.

\bibitem[{\citenamefont{Tame et~al.}(2005)\citenamefont{Tame, Paternostro, Kim,
  and Vedral}}]{05-tam-qua}
\bibinfo{author}{\bibfnamefont{M.~S.} \bibnamefont{Tame}},
  \bibinfo{author}{\bibfnamefont{M.}~\bibnamefont{Paternostro}},
  \bibinfo{author}{\bibfnamefont{M.~S.} \bibnamefont{Kim}}, \bibnamefont{and}
  \bibinfo{author}{\bibfnamefont{V.}~\bibnamefont{Vedral}},
  \bibinfo{journal}{Phys. Rev. A.} \textbf{\bibinfo{volume}{72}},
  \bibinfo{pages}{012319} (\bibinfo{year}{2005}).

\bibitem[{\citenamefont{Raussendorf et~al.}(2003)\citenamefont{Raussendorf,
  Browne, and Briegel}}]{03-rau-mea}
\bibinfo{author}{\bibfnamefont{R.}~\bibnamefont{Raussendorf}},
  \bibinfo{author}{\bibfnamefont{D.~E.} \bibnamefont{Browne}},
  \bibnamefont{and} \bibinfo{author}{\bibfnamefont{H.~J.}
  \bibnamefont{Briegel}}, \bibinfo{journal}{Phys. Rev. A}
  \textbf{\bibinfo{volume}{68}}, \bibinfo{pages}{022312}
  (\bibinfo{year}{2003}).

\bibitem[{\citenamefont{Cinelli et~al.}(2005)\citenamefont{Cinelli, Barbieri,
  Perris, Mataloni, and {De Martini}}}]{05-cin-all}
\bibinfo{author}{\bibfnamefont{C.}~\bibnamefont{Cinelli}},
  \bibinfo{author}{\bibfnamefont{M.}~\bibnamefont{Barbieri}},
  \bibinfo{author}{\bibfnamefont{R.}~\bibnamefont{Perris}},
  \bibinfo{author}{\bibfnamefont{P.}~\bibnamefont{Mataloni}}, \bibnamefont{and}
  \bibinfo{author}{\bibfnamefont{F.}~\bibnamefont{{De Martini}}},
  \bibinfo{journal}{Phys. Rev. Lett.} \textbf{\bibinfo{volume}{95}},
  \bibinfo{pages}{240405} (\bibinfo{year}{2005}).

\bibitem[{\citenamefont{Gottensman}(1997)}]{97-got-sta}
\bibinfo{author}{\bibfnamefont{D.}~\bibnamefont{Gottensman}}, Ph.D. thesis,
  \bibinfo{school}{CalTech, Pasadena} (\bibinfo{year}{1997}).

\bibitem[{\citenamefont{Hein et~al.}(2006)\citenamefont{Hein, D{\"u}r, Eisert,
  Raussendorf, den Nest, and Briegel}}]{06-hei-ent}
\bibinfo{author}{\bibfnamefont{M.}~\bibnamefont{Hein}},
  \bibinfo{author}{\bibfnamefont{W.}~\bibnamefont{D{\"u}r}},
  \bibinfo{author}{\bibfnamefont{J.}~\bibnamefont{Eisert}},
  \bibinfo{author}{\bibfnamefont{R.}~\bibnamefont{Raussendorf}},
  \bibinfo{author}{\bibfnamefont{M.~V.} \bibnamefont{den Nest}},
  \bibnamefont{and} \bibinfo{author}{\bibfnamefont{H.-J.}
  \bibnamefont{Briegel}}, in \emph{\bibinfo{booktitle}{Quantum computers,
  algorithms and chaos}}, edited by
  \bibinfo{editor}{\bibfnamefont{P.}~\bibnamefont{Zoller}},
  \bibinfo{editor}{\bibfnamefont{G.}~\bibnamefont{Casati}},
  \bibinfo{editor}{\bibfnamefont{D.}~\bibnamefont{Shepelyansky}},
  \bibnamefont{and} \bibinfo{editor}{\bibfnamefont{G.}~\bibnamefont{Benenti}}
  (\bibinfo{year}{2006}), International School of Physics Enrico Fermi
  (Varenna, Italy),
  \eprint{{\texttt{\href{http://www.arxiv.org/pdf/quant-ph/0602096}{[quant-ph/%
0602096]}}}}.

\bibitem[{\citenamefont{Grover}(1997{\natexlab{a}})}]{97-gro-qua}
\bibinfo{author}{\bibfnamefont{L.~K.} \bibnamefont{Grover}},
  \bibinfo{journal}{Phys. Rev. Lett.} \textbf{\bibinfo{volume}{79}},
  \bibinfo{pages}{325} (\bibinfo{year}{1997}{\natexlab{a}}).

\bibitem[{\citenamefont{Grover}(1997{\natexlab{b}})}]{97-gro-qua2}
\bibinfo{author}{\bibfnamefont{L.~K.} \bibnamefont{Grover}},
  \bibinfo{journal}{Phys. Rev. Lett.} \textbf{\bibinfo{volume}{79}},
  \bibinfo{pages}{4709} (\bibinfo{year}{1997}{\natexlab{b}}).

\bibitem[{\citenamefont{Tame et~al.}(2007)\citenamefont{Tame, Prevedel,
  Paternostro, Bohi, Kim, and Zeilinger}}]{07-tam-exp}
\bibinfo{author}{\bibfnamefont{M.~S.} \bibnamefont{Tame}},
  \bibinfo{author}{\bibfnamefont{R.}~\bibnamefont{Prevedel}},
  \bibinfo{author}{\bibfnamefont{M.}~\bibnamefont{Paternostro}},
  \bibinfo{author}{\bibfnamefont{P.}~\bibnamefont{Bohi}},
  \bibinfo{author}{\bibfnamefont{M.~S.} \bibnamefont{Kim}}, \bibnamefont{and}
  \bibinfo{author}{\bibfnamefont{A.}~\bibnamefont{Zeilinger}},
  \bibinfo{journal}{Phys. Rev. Lett.} \textbf{\bibinfo{volume}{98}},
  \bibinfo{pages}{140501} (\bibinfo{year}{2007}).

\end{thebibliography}
\end{document}